\documentclass{iopart}

\usepackage{iopams}
\usepackage{amssymb}
\usepackage{graphicx}
\usepackage{mathrsfs}

\def\RnR{rock'n'roller}

\def\half{\textstyle\frac{1}{2}}
\def\iprime{i^\prime}
\def\jprime{j^\prime}
\def\kprime{k^\prime}
\def\IA{I_1} \def\bIA{\bi{I_1}}
\def\IB{I_2} \def\bIB{\bi{I_2}}
\def\IC{I_3} 
\def\bvdot{\mathbf{\dot\bi{v}}}
\def\bqdot{\mathbf{\dot\bi{q}}}
\def\bomdot{\mathbf{\dot{\bomega}}}
\def\bthdot{\mathbf{\dot{\btheta}}}
\def\bthddot{\mathbf{\ddot{\btheta}}}
\def\cross{\boldsymbol{\times}}
\def\cphi{c_\phi}
\def\sphi{s_\phi}

\begin{document}

\title{Precession and recession of the \RnR}

\author{Peter Lynch and Miguel D Bustamante}

\address{School of Mathematical Sciences,
        UCD, Belfield, Dublin 4, Ireland}
\ead{\mailto{Peter.Lynch@ucd.ie},
    \mailto{Miguel.Bustamante@ucd.ie}}




\begin{abstract}
We study the dynamics of a spherical rigid body that rocks and
rolls on a plane under the effect of gravity. 
The distribution of mass is non-uniform and the centre of mass
does not coincide with the geometric centre.
The symmetric case, with
moments of inertia $\IA=\IB<\IC$, is integrable and the motion
is completely regular. Three known conservation laws are the
total energy $E$, Jellett's quantity $Q_J$ and Routh's
quantity $Q_R$. When the inertial symmetry $\IA=\IB$ is
broken, even slightly, the character of the solutions is
profoundly changed and new types of motion become possible. We
derive the equations governing the general motion and present
analytical and numerical evidence of the recession, or
reversal of precession, that has been observed in physical
experiments. We present an analysis of recession in terms of
critical lines dividing the $(Q_R,Q_J)$ plane into four
dynamically disjoint zones. We prove that recession implies
the lack of conservation of Jellett's and Routh's quantities,
by identifying individual reversals as crossings of the orbit
$(Q_R(t),Q_J(t))$ through the critical lines. Consequently, a
method is found to produce a large number of initial
conditions so that the system will exhibit recession.

\end{abstract}


\section{Introduction}
\label{sec:Introduction}

We investigate the dynamics of a spherical rigid body rolling
on a plane. The distribution of mass is non-uniform, so that
the centre of mass does not coincide with the geometric
centre. However, the line joining the mass centre and
geometric centre is assumed to be a principal axis. We denote
the principal moments of inertia by $\IA$, $\IB$ and $\IC$,
and assume that $\IA\le \IB < \IC$. The symmetric case, when
$\IA=\IB$, was first studied by Routh \cite{Routh05}, and in
this case the body is called Routh's Sphere. There are three
constants of motion and the system is integrable. In the
asymmetric case, $\IA\ne \IB$, the system is no longer
integrable. We find that even a small degree of asymmetry has
a dramatic effect on the motion of the body.

The equations of the symmetric loaded sphere are identical to
those governing the motion of the tippe-top, which has been
studied extensively (see \cite{GrayNickel00} for a
comprehensive reference list). However, in the case of the
tippe-top, the angular momentum about the principal axis with
maximum moment of inertia is large, and sliding friction plays
a key role. In the case under consideration here, we are
interested in solutions where the angular velocity remains
moderate and there is pure rolling contact. There are two
characteristic modes of behaviour: pure rocking motion in a
vertical plane, and pure circular rolling motion. The general
motion has aspects of both these special cases, which leads us
to name the body the {\em \RnR}.

\begin{figure}
\begin{center}
\includegraphics[scale=0.50]{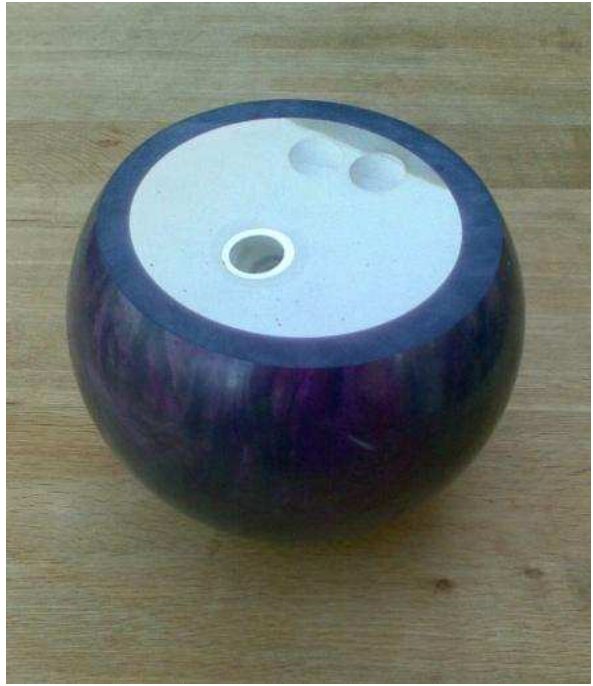}
\caption{The physical \RnR, constructed by slicing off a polar
cap from a standard bowling ball.
The polar angle is $\Theta\approx 53^\circ$.}
\label{fig:bowlingball}
\end{center}
\end{figure}

This investigation arose from the observation of the
oscillations of a glass candle holder, spherical in form with
an opening at the top. For a more systematic study, we
constructed a larger and more massive body by removing a polar
cap from a bowling ball to produce a truncated sphere
(figure~\ref{fig:bowlingball}). As long as the tilting angle
is such that the geometric centre is vertically above the
contact point, the dynamics are equivalent to those of a
loaded sphere. It was found that when the ball was tilted over
to an angle of about $130^\circ$, it rocked back and forth but
also precessed through an azimuthal angle that alternately
increased and decreased. This unexpected and surprising {\em
recession}, or reversal of precession, demanded an explanation
in terms of dynamics.

We will show that for a symmetric loaded sphere reversal of
the precession is impossible. This raises the question: what
factor is missing from our dynamical model? We rule out
sliding friction, since the motion is gentle with no evidence
of slipping. Random perturbations, due to the imperfect shape
of the ball or irregularities of the underlying surface, were
not considered as a likely cause of the behaviour, as
experiments indicated that the recession was quite a robust
feature of the motion.

Although bowling balls are manufactured to high tolerence, and
deviations from perfect sphericity must be very small, slight
anomalies in the mass distribution are unavoidable. Moreover,
the recesses in the physical body remaining from the finger
holes introduce some asymmetry (figure~\ref{fig:bowlingball}).
We were thus led to study the dynamics when the inertial
symmetry $\IA=\IB$ is broken. We find that even a minute
deviation from symmetry changes the behaviour of the numerical
solution profoundly. Of the three quantities conserved in the
symmetric case     
(total energy $E$, Jellett's quantity $Q_J$ and Routh's
quantity $Q_R$), only the energy remains invariant when
$\IA\ne\IB$. We derive the equations governing the general
motion and present analytical and numerical evidence of
recession. We base our analysis on the existence of critical
lines dividing the $(Q_R,Q_J)$ plane into four dynamically
disjoint zones. We prove that recession implies the lack of
conservation of Jellett's and Routh's quantities, by
identifying individual reversals as crossings of the orbit
$(Q_R(t),Q_J(t))$ through the critical lines. This leads to a
method of defining initial conditions for which the system will
exhibit recession.

\begin{figure}
\begin{center}
\includegraphics[scale=0.15]{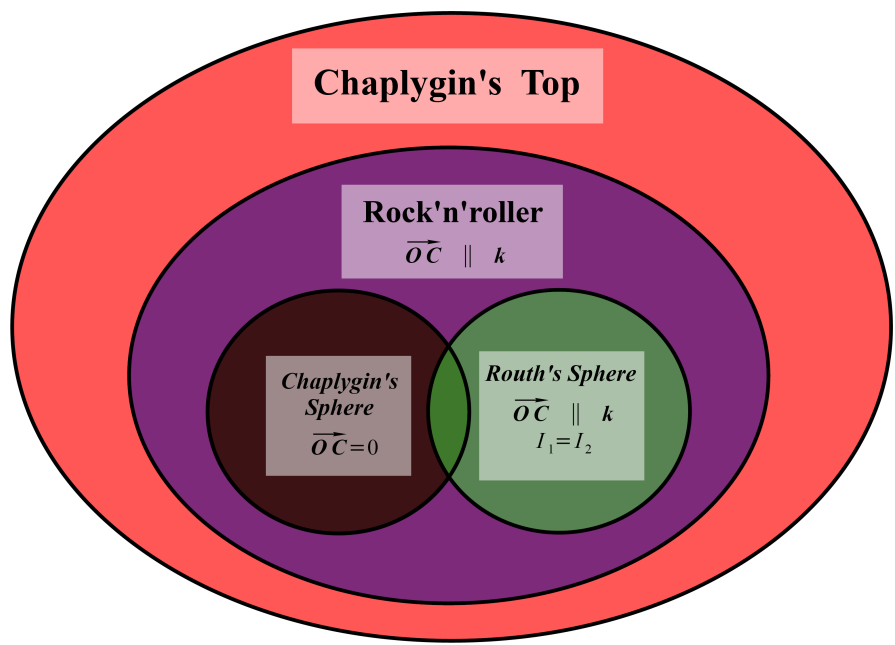}
\caption{Hierarchy of loaded spheres. The vector
$\overrightarrow{OC}$ is from the mass centre $O$ to the
geometric centre $C$, and $\mathbf{k}$ is the unit vector
along the $\IC$-axis. See text for details.}
\label{fig:LoadedSpheres}
\end{center}
\end{figure}

The \RnR\ is one of a hierarchy of loaded spheres. For the
most general case, the vector $\overrightarrow{OC}$ from the
mass centre $O$ to the geometric centre $C$ does not lie on a
principal axis, and all moments of inertia are distinct. This
is called Chaplygin's Top \cite{Chaplygin03}. For the \RnR,
the geometric centre lies on a principal axis and
$\overrightarrow{{OC}}$ is parallel to $\mathbf{k}$, the unit
vector along the $\IC$-axis. Routh's Sphere is the special
case of this with $\IA=\IB$ and Chaplygin's Sphere the special
case where the mass centre and geometric centre coincide. The
hierarchy is illustrated in figure~\ref{fig:LoadedSpheres}.
For recent discussions, see \cite{Holm08ab,
      Duistermaat04,
      Cushman98,
      Kilin01,
      ShenSchneiderBloch03,
      BorisovMamaev02,
      Schneider02}

See an animation of the \RnR\ that exhibits precession and recession in  
\verb1http://mathsci.ucd.ie/~plynch/RnR/RnR_movie.gif1.
We produced this movie from a \emph{Mathematica} simulation code 
of the equations in the asymmetric case $\IA\ne\IB$,
corresponding to the initial conditions described in Figure~\ref{fig:reversals(QR,QJ)2}.


\section{Symmetric Body ($\bIA=\bIB$): The Dynamical Equations}
\label{sec:SymmDynamics}

We consider a body, spherical in shape with unit mass and unit
radius, whose mass distribution is non-uniform but symmetric
about some line through the centre. We assume that the centre
of mass is off-set a distance $a$ from the geometric centre
and that the moments of inertia perpendicular to and along the
symmetry axis are $\IA$ and $\IC$, with $\IA(=\IB)<\IC$. All
the parameters are determined once the angle of the polar cap
that is removed is known (see Appendix \S A.1). In an inertial
frame of reference, the equations governing the dynamics of
the body are
\begin{equation}
\frac{\rmd\bi{v}}{\rmd t} = \bi{F} \label{eq:vstatic}
\end{equation}
where $\bi{v}$ is the velocity of the centre of mass in the
absolute frame and $\bi{F}$ the total force acting on the
body; and
\begin{equation}
\frac{\rmd\bi{L}}{\rmd t} = \bi{G} \label{eq:Lstatic}
\end{equation}
where $\bi{L}$ is the intrinsic angular momentum and $\bi{G}$
the total moment about the centre of mass.

\begin{figure}
\begin{center}
\includegraphics[scale=0.60]{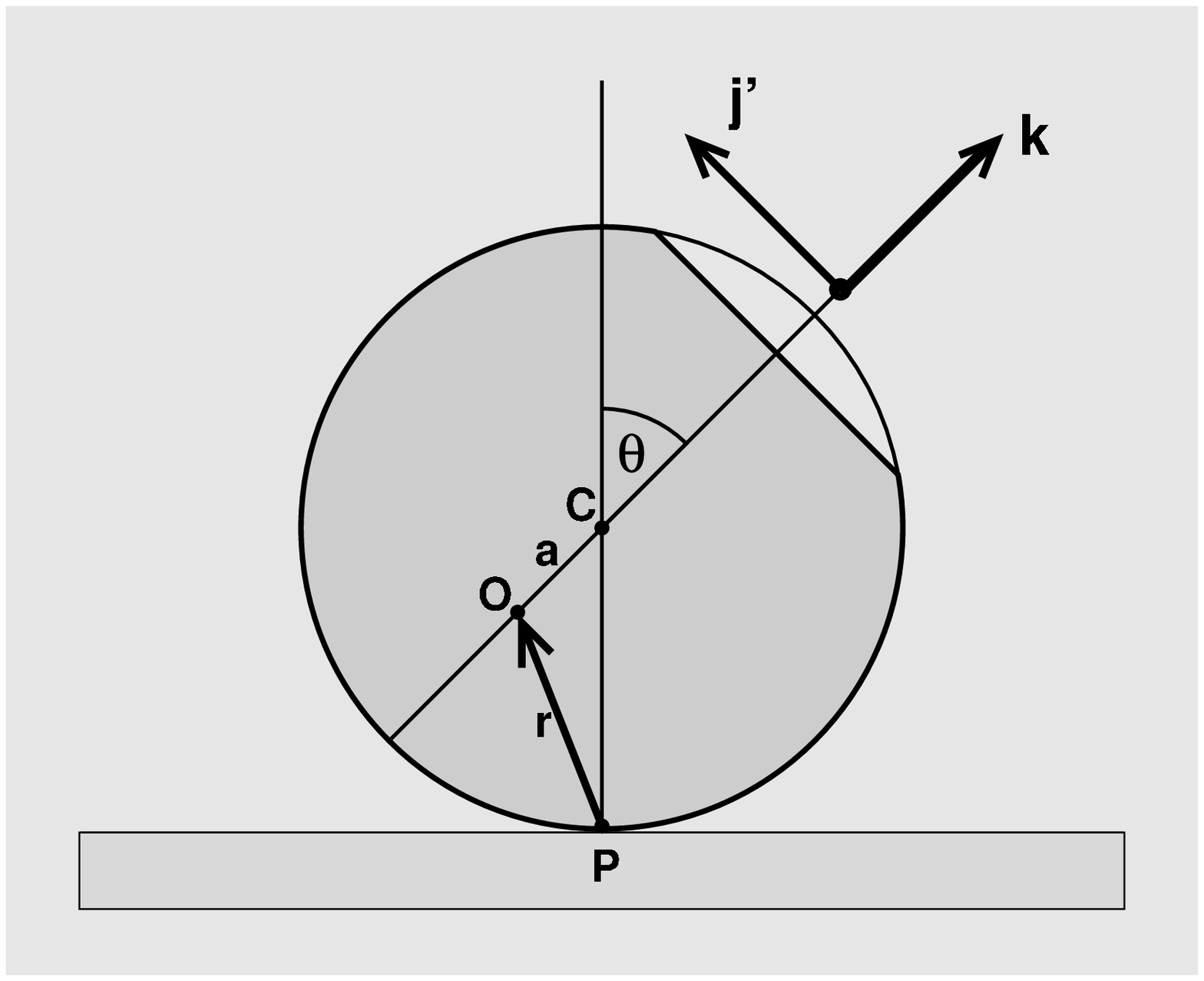}
\caption{The intermediate coordinate frame used to study
         Routh's Sphere.}
\label{fig:coordinates}
\end{center}
\end{figure}

The derivation in this section is similar to that in
\cite{UedaSasakiWatanabe05}. We consider a rotating frame of
reference, with unit triad
$(\bi{\iprime},\bi{\jprime},\bi{\kprime})$ whose origin moves
with the centre of mass of the body.  The vector
$\bi{\kprime}$ is aligned with the axis of symmetry of the
body and $\bi{\jprime}$ is in the same vertical plane as
$\bi{\kprime}$ (see figure~\ref{fig:coordinates}). Then
$\bi{\iprime}$ is horizontal and perpendicular to the plane of
the figure, pointing inward. We use primes for this
intermediate frame to distinguish it from the body frame that
will be introduced in \S\ref{sec:AsymDynamics} below.

The angular velocity of the body, expressed in the
intermediate frame, is
\[
{\bomega} = \omega_1^\prime\bi{\iprime} +
\omega_2^\prime\bi{\jprime} + \omega_3^\prime\bi{\kprime}
\]
Although this frame is not fixed in the body, it forms a set
of principal axes in the symmetric case, $\IA=\IB$, and the
angular momentum is given by
\[
\bi{L} = \IA\omega_1^\prime\bi{\iprime} +
\IA\omega_2^\prime\bi{\jprime} +
\IC\omega_3^\prime\bi{\kprime} \,.
\]
We denote the angular velocity of the frame itself by
$\bOmega$ and note that
\begin{equation}
\bOmega
 = \dot\theta\bi{\iprime}+\dot\phi\bi{K}
 = \dot\theta\bi{\iprime}+s\dot\phi\bi{\jprime}+c\dot\phi\bi{\kprime}
 = (\Omega_1,\Omega_2,\Omega_3)
\label{eq:Omega}
\end{equation}
where $s=\sin\theta$ and $c=\cos\theta$ and $\bi{K}$ is a unit
vertical vector.  The Euler angles $(\theta,\phi,\psi)$ are
related to the components of angular velocity by
\begin{equation}
\omega_1^\prime =   \dot\theta          \,,\qquad
\omega_2^\prime = s \dot\phi            \,,\qquad
\omega_3^\prime = c \dot\phi + \dot\psi \,.
\label{eq:angledot}
\end{equation}
Definitions are standard, and may be found in
\cite{MontaldiRatiu05, SyngeGriffith59, Whittaker37}. For a
list of the principal symbols used in this study, see
Table~\ref{tab:Notation}.

%
%

\begin{table}
\caption{Principal symbols used in this study}
\label{tab:Notation}
\medskip
\begin{center}
\begin{tabular}{ l l }
\hline\hline
Symbol  &   Meaning  \\
\hline
%
%
$\bi F$        &   Total forcing in Newton's equation   \\
$\bi G$        &   Total moment in Newton's equation   \\
$\IA$, $\IB$, $\IC$   &   Principal moments of inertia of body \\
$\bi K$        &   Unit vertical vector \\
$\bi L$        &    Angular momentum of body about centre of mass \\
$\cal L$       &    Lagrangian function \\
$Q_J$    &   Jellett's quantity, constant in symmetric case \\
$Q_R$    &   Routh's quantity, constant in symmetric case \\
$\bi R$  &   Force of reaction at contact point \\
$T$      &   Total kinetic energy \\
$V$      &   Potential energy \\
$\bi V$  &   Velocity of centre of mass in the space frame \\
$\bi W$  &   Force due to gravity (weight) \\
%
%
$a$            & Distance from geometric centre to
                 centre of mass \\
$c$            & Cosine of tilting angle, $c=\cos\theta$  \\
$c_\phi$       & Cosine of azimuthal angle, $c_\phi=\cos\phi$  \\
$d$            & Cosine of polar angle, $d=\cos\Theta$ \\
$f$            & Projection of vertical radius on
        $\bi k$-axis, $f=\cos\theta - a$ \\
$g$            & Acceleration of gravity \\
$h$            & Height of centre of mass, $h = 1-a\cos\theta$ \\
$\bi{i, j, k}$ &
    Principal unit orthogonal triad in body coordinates \\
$\bi{\iprime, \jprime, \kprime}$ &
    Principal unit orthogonal triad in body coordinates \\
$\bi r$        & Moment vector, from contact point to mass centre \\
$s$            & Sine of tilting angle, $s=\sin\theta$  \\
$s_\phi$       & Sine of azimuthal angle, $s_\phi=\sin\phi$  \\
$t$            & Time  \\
$\bi v$        &   Velocity of centre of mass in the body frame \\
$v_1$, $v_2$, $v_3$   &
    Components of $\bi v$ in body coordinates \\
%
%
$\Theta$      & Co-latitude of polar cap removed
                to construct the \RnR.  \\
$\bUpsilon$    &   Rotation matrix  \\
$\Phi$         &   Azimuthal angle spanned by solution,
                   $\Phi=\phi_{\rm max}-\phi_{\rm min}$  \\
$\bOmega$      &   Angular velocity of intermediate frame  \\
%
%
$\epsilon$     &   Asymmetry parameter, $\epsilon=(\IB-\IA)/\IA$ \\
$\theta, \phi, \psi$  &
      Euler angles (tilting, azimuth and spinning angles) \\
$\mu_k$        &   Lagrange multipliers \\
$\rho$         &
      Measure quantity, $\rho=[\IC+s^2+(\IC/\IA)f^2]^{-1/2}$ \\
$\sigma$       & Sine of spining angle, $\sigma=\sin\psi$  \\
$\tau$         & Period of rocking motion  \\
$\chi$         & Cosine of spinning angle, $\chi=\cos\psi$  \\
$\bomega$      &   Angular velocity of body   \\
$\omega_1^\prime$, $\omega_2^\prime$, $\omega_3^\prime$     &
    Components of $\bomega$ in intermediate coordinates \\
$\omega_1$, $\omega_2$, $\omega_3$     &
    Components of $\bomega$ in body coordinates \\
\hline\hline
\end{tabular}
\end{center}
\end{table}


\subsection{Equations in the intermediate frame}
\label{sec:intermediateframe}

In the moving frame, the equations (\ref{eq:vstatic}) and
(\ref{eq:Lstatic}) become
\begin{equation}
\frac{\rmd\bi{v}}{\rmd t} + \bOmega\bi{\cross v} = \bi{F}
\label{eq:vrot}
\end{equation}
and
\begin{equation}
\frac{\rmd\bi{L}}{\rmd t} + \bOmega\bi{\cross L} = \bi{G}
\label{eq:Lrot}
\end{equation}
Expanding these in components in the $\bi{i^\prime j^\prime
k^\prime }$-frame, we get
\begin{eqnarray}
\dot v_1^\prime +
\Omega_2 v_3^\prime - \Omega_3 v_2^\prime &=& F_1 \nonumber \\
\dot v_2^\prime + \Omega_3 v_1^\prime - \Omega_1 v_3^\prime
&=& F_2
\label{eq:Translation} \\
\dot v_3^\prime + \Omega_1 v_2^\prime - \Omega_2 v_1^\prime
&=& F_3 \nonumber
\end{eqnarray}
for momentum.  The angular momentum equations become
\begin{eqnarray}
\IA\dot\omega_1^\prime
+ \IC\Omega_2\omega_3^\prime - \IA\Omega_3\omega_2^\prime &=& G_1 \nonumber \\
\IB\dot\omega_2^\prime + \IA\Omega_3\omega_1^\prime -
\IC\Omega_1\omega_3^\prime &=& G_2
\label{eq:Rotation} \\
\IC\dot\omega_3^\prime
&=& G_3 \nonumber
\end{eqnarray}
Eqns.~(\ref{eq:Translation}) and (\ref{eq:Rotation}) are
identical to (12.412) in \cite{SyngeGriffith59} (with
$\IA=\IB$).

The forces acting on the body are gravity $\bi{W}=(0,-gs,-gc)$
and the force of reaction $\bi{R}=(R_1,R_2,R_3)$:
\[
\bi{   F  =  W + R   }
\]
Defining $f=c-a$, the vector from the point of contact $P$ to
the centre of mass $O$ is $\bi{r}=(0,s,f)$ (see
figure~\ref{fig:coordinates}). Then the total moment about $O$
is given by $\bi{G} = -\bi{r\cross R}$. The \emph{constraint
of no slipping} at the contact point requires that the body is
instantaneously rotating about this point. Thus,
\begin{equation}
\bi{v} = \bi{\bomega\cross r}
 = (f\omega_2^\prime-s\omega_3^\prime,-f\omega_1^\prime,s\omega_1^\prime) \,.
\label{eq:constraint}
\end{equation}
The reactive forces may be eliminated by combining the angular
momentum equation (\ref{eq:Lrot}) with the vector product of
$\bi r$ and the momentum equation (\ref{eq:vrot}).  The
velocity $\bi{v}$ may be expressed in terms of the rotation
$\bomega$ by means of the constraint (\ref{eq:constraint}).
We then obtain three equations for $\omega_1$, $\omega_2$ and
$\omega_3$:
\begin{equation}
\left[\matrix{ \IA+s^2+f^2 &    0      &    0     \cr
                     0     &  \IA+f^2  &   -fs    \cr
                     0     &    -fs    &  \IC+s^2  } \right]
\pmatrix{\dot\omega_1^\prime \cr \dot\omega_2^\prime \cr
\dot\omega_3^\prime} = \pmatrix{  P_1  \cr  P_2  \cr P_3}
\label{eq:omegasdot}
\end{equation}
where $P_1$, $P_2$ and $P_3$ depend on the angles and angular
velocities.  Full details of the derivation are presented in
the Appendix (\S A.2). The rates of change of the angular
variables follow from (\ref{eq:angledot}):
\begin{equation}
\dot\theta = \omega_1^\prime   \,, \qquad \dot\phi  =
\omega_2^\prime/s  \,, \qquad \dot\psi = \omega_3^\prime -
(c/s) \omega_2^\prime  \,. \label{eq:anglesdot}
\end{equation}
We now have six equations (\ref{eq:omegasdot}) and
(\ref{eq:anglesdot}) for the six variables
$\{\theta,\phi,\psi,\omega_1^\prime,\omega_2^\prime,\omega_3^\prime\}$.

\subsection{Special solutions}

\subsubsection*{Pure Rocking.}
For pure rocking motion, with no change of azimuthal angle and
no rotation about the axis of symmetry, we have $\phi=\psi=0$
and so $\omega_2^\prime=\omega_3^\prime=0$. Then the system
reduces to a single equation for the tilting angle $\theta$:
\begin{equation}
\ddot\theta + \left[\displaystyle{
\frac{(g+\dot\theta^2)a}{\IA+f^2+s^2} }\right]\sin\theta = 0
\label{eq:purerock}
\end{equation}
For small amplitude $\theta\ll 1$, and assuming $a\ll 1$, this
becomes
\begin{equation}
\ddot\theta + \left[\displaystyle{ \frac{ga}{\IA+1} } \right]
\theta = 0 \,, \label{eq:smallamprock}
\end{equation}
the equation for simple harmonic oscillations.

\subsubsection*{Pure Rolling.}
For the case of pure circular rolling motion we have
\[
\dot\theta = 0 \,, \qquad \dot\phi = \mbox{\rm constant} \,,
\qquad \dot\psi = \mbox{\rm constant}
\]
so that $\Omega_1=\omega_1^\prime=0$ and $\omega_2^\prime$ and
$\omega_3^\prime$ are constants. It follows immediately that
$P_2=P_3=0$ (see (\ref{eq:PPP})). The requirement that
$\theta=\theta_0$, constant, implies $P_1=0$, which yields a
relationship between $\omega_2^\prime$ and $\omega_3^\prime$:
\begin{equation}
\omega_3^\prime = \frac{(\IA\cot\theta_0 +
mh_0f_0\csc\theta_0)\omega_2^{\prime2}
                          -ga\sin\theta_0}    {(\IC+h_0)\omega_2^\prime}
\label{eq:om2om3}
\end{equation}
where $f_0=\cos\theta_0-a$ and $h_0=1-a\cos\theta_0$ are
constants.
If we start with $\omega_2^\prime$ and $\omega_3^\prime$
related by (\ref{eq:om2om3}) and $\theta$ slightly perturbed
from $\theta_0$, motion with nutation about $\theta_0$
results.

\subsection{Constants of motion and general solution}
\label{sec:Constants}

We consider the case of a perfectly rough contact, with
rolling motion.  Given that there are two symmetries in the
problem, invariance under addition of arbitrary constants to
either $\phi$ or $\psi$, we might expect two invariants in
addition to the total energy.  For general initial conditions,
there are three constants of integration. They are the total
energy, Jellett's constant and Routh's constant (see
\cite{GrayNickel00} for a complete derivation of these
constants).

The kinetic energy is the sum of translational and rotational
components:
\[
T = \half[v_1^{\prime2}+v_2^{\prime2}+v_3^{\prime2}] +
\half[\IA\omega_1^{\prime2}+\IB\omega_2^{\prime2}+\IC\omega_3^{\prime2}]
\label{eq:kinetic}
\]
and the potential energy is
\[
V = ga(1-\cos\theta) \,.
\]
Then, since there is no dissipation, the total energy
\begin{equation}
E = T + V  \,. \label{eq:energy}
\end{equation}
is conserved.  Jellett's constant is the scalar product of the
angular momentum and the vector joining the point of contact
to the centre of mass:
\begin{equation}
Q_J = \bi{L\cdot r} = \IA s\,\omega_2^\prime + \IC
f\,\omega_3^\prime \,, \label{eq:jellett}
\end{equation}
and Routh's constant, more difficult to interpret physically,
is
\begin{equation}
Q_R = \frac{\omega_3^{\prime}}{\rho} \label{eq:routh}
\end{equation}
where, following \cite{BorisovMamaev02}, we define the measure
\begin{equation}
\rho(\theta) = 1/\sqrt{\IC + s^2 + (\IC/\IA)f^2} \,.
\label{eq:measure}
\end{equation}
Notice that our definition of Routh's constant differs from
the usual quadratic function of $\omega_3^{\prime}$, in
\cite{Routh05}, \cite{GrayNickel00} and elsewhere,

An interesting historical discussion of these constants may be
found in \cite{GrayNickel00}. Note that the constancy of $Q_R$
implies conservation of the \emph{sign} of $\omega_3^\prime$:
since the measure $\rho$ is positive definite,
$\omega_3^\prime$ cannot pass through zero. For the tippe-top,
this precludes the tipping phenomenon for the case of rolling
motion.

From the equations (\ref{eq:anglesdot}) determining the rates
of change of the angles, we can solve explicitly for
$\dot\phi$ and $\dot\psi$ in terms of $\theta$ and Jellett's
and Routh's constants:
\begin{eqnarray}
\label{eq:phidotCJQR} \dot\phi &=& V_{\phi}(\theta,Q_J,Q_R)
\equiv
\frac{1}{\IA s^2}\biggl[Q_J-\rho f \IC Q_R \biggr] \,, \\
\label{eq:psidotCJQR} \dot\psi &=& V_{\psi}(\theta,Q_J,Q_R)
\equiv -\frac{1}{\IA s^2}\biggl[c Q_J-\rho(c f\IC+\IA
s^2)Q_R\biggr] \,.
\end{eqnarray}
Since $Q_J$ and $Q_R$ are constants, the rates of change
$\dot\phi$ and $\dot\psi$ are determined as single-valued
functions of the angle $\theta$. We will show in the next
section that recession, or reversal of precession, implies in
particular that $\dot\phi$ and $\dot\psi$ at a given angle
$\theta$ systematically change their sign as time evolves.
Therefore, in the symmetric case it is impossible to have
recession for Routh's Sphere.

We can use the constants of motion to reduce the system to a
single equation for the tilting angle $\theta$. We use Routh's
constant (\ref{eq:routh}) to obtain $\omega_3^\prime(\theta)$.
Then Jellett's constant (\ref{eq:jellett}) gives
$\omega_2^\prime(\theta)$. Finally, the energy
(\ref{eq:energy}) gives an expression for
$\omega_1^\prime(\theta)$, yielding an equation of the form
\begin{equation}
\dot\theta^2 = F(\theta) \,, \label{eq:thetadotsq}
\end{equation}
which may be integrated to obtain $\theta(t)$. As a result,
the system can be explicitly integrated. However, we will not
derive explicit analytical expressions for $F(\theta)$ and
$\theta(t)$. The reader is referred to \cite{GrayNickel00} for
a more explicit treatment; see also \cite{Cushman98},
\cite{Schneider02}. We see that the evolution of $\theta(t)$
obtained from (\ref{eq:thetadotsq}) gives the rocking
component of the motion, while the evolution of $\phi(t)$ and
$\psi(t)$, obtained from (\ref{eq:phidotCJQR}) and
(\ref{eq:psidotCJQR}), give the rolling and spinning
components of the motion.

\subsection{Precession of the Rocking Motion}

The generic motion of the symmetric body is quasi-periodic. On
the one hand there is a period $\tau$ of the rocking motion,
determined by the equation
$$
\tau = 4 \int_{\theta_{N}}^{\theta_X}
\frac{\rmd\theta}{\sqrt{F(\theta)}} \,,
$$
where $0\le\theta_{N}\le\theta_{X}\le\pi$, and $\theta_{N}$
and $\theta_{X}$ are the {\em turning points} where
$F(\theta_{N})=0$ and $F(\theta_{X})=0$. On the other hand,
the rolling motion during this period can be computed by
integrating the rates of change of the angles $\phi$ and
$\psi$ from (\ref{eq:phidotCJQR}) and (\ref{eq:psidotCJQR}):
\begin{equation}
\label{eq:Delta_phi} \Delta \phi = \int_{0}^\tau
V_{\phi}(\theta(t),Q_J,Q_R) \,\rmd t
 = 4 \int_{\theta_{N}}^{\theta_X}
\displaystyle{
\frac{V_{\phi}(\theta,Q_J,Q_R)}{\sqrt{F(\theta)}} }\,{\rmd
\theta} \,,
\end{equation}
with an analogous formula for the angle $\Delta \psi$.
Generically, $\Delta \phi$ is not commensurate with $2 \pi$;
this implies the quasi-periodicity of the precessing motion.
As a consequence of quasi-periodicity, the projection of the
trajectory onto the $\theta$-$\phi$-plane densely covers a
two-dimensional region.

In order to quantify the precession, we distinguish two
angles, the \emph{full azimuthal angle} $\phi$ and the
\emph{visible angle} $\phi\,(\mathrm{mod}\,2\pi)$, which is
the angle that is seen by an observer in the space frame. (We
will occasionally use the \emph{visible half-angle}
$\phi\,(\mathrm{mod}\,\pi)$, which gives more illustrative
plots in the case of the (asymmetric) \RnR). Correspondingly,
there will be two types of precession angle: $\Delta\phi$, the
\emph{full precession angle} defined by (\ref{eq:Delta_phi}),
and $\Delta\phi\,(\mathrm{mod}\,2\pi)$, the \emph{visible
precession angle}.

\subsection{Qualitative analysis of the precession. Criticality}
\label{sec:criticality}

We now estimate the precession angles $\Delta \phi$ and
$\Delta\phi\,(\mathrm{mod}\,2\pi)$ from (\ref{eq:Delta_phi}).
Heuristically, the main contribution comes from the regions
near the turning points, where $F(\theta)=0$. The relative
contributions at $\theta_N$ and $\theta_X$ will be determined
by the magnitude and sign of $V_\phi(\theta,Q_J,Q_R)$ at the
turning points. It is therefore useful to study separately the
behaviour of $V_\phi(\theta,Q_J,Q_R)$ for $\theta$ in each of
the asymptotic regions $\theta\approx 0$ and
$\theta\approx\pi$:

\begin{itemize}
\item Asymptotic region $\theta \approx 0$.
A Laurent expansion of (\ref{eq:phidotCJQR}) gives
\begin{eqnarray}
\label{eq:phidotCJQR th=0} \left. V_{\phi}(\theta,Q_J,Q_R)
\right|_{\theta \approx 0} &=& \frac{1}{{\IA}\,\theta^2}\,
\left({Q_J}-Q_{J,0}^{\mathrm{crit}}\right) + \mathcal{O}(1)\,,
\end{eqnarray}
 where we define the `critical Jellett quantity at $\theta = 0$' as:
 \begin{equation}
 \label{eq:QJcrit_0}
 Q_{J,0}^{\mathrm{crit}} \equiv \rho_0 (1-a)\IC Q_R \,,
 \end{equation}
with $\rho_0=1/\sqrt{\IC+(\IC/\IA)(1-a)^2}$.
\item Asymptotic region $\theta \approx \pi$.
A Laurent expansion of (\ref{eq:phidotCJQR}) gives
\begin{eqnarray}
\label{eq:phidotCJQR th=pi} \left. V_{\phi}(\theta,Q_J,Q_R)
\right|_{\theta \approx \pi} &=&
\frac{1}{{\IA}\,(\pi-\theta)^2}\,
\left({Q_J}-Q_{J,\pi}^{\mathrm{crit}}\right) +
\mathcal{O}(1)\,,
\end{eqnarray}
 where we define the `critical Jellett quantity at $\theta =
\pi$' as:
 \begin{equation}
 \label{eq:QJcrit_pi}
 Q_{J,\pi}^{\mathrm{crit}} \equiv -\rho_{\pi} (1+a)\IC Q_R \,.
\end{equation}
with $\rho_{\pi}=1/\sqrt{\IC+(\IC/\IA)(1+a)^2}$.
\item Monotonicity property.
The factor $[Q_J - \rho f\IC Q_R]$ appearing on the right-hand
side of (\ref{eq:phidotCJQR}) is a monotonic function of the
angle $\theta \in [0,\pi]$.
The proof of this is straightforward.
\end{itemize}
From the above asymptotic expansions, we conclude that, in the
space of initial conditions parameterised by $(Q_R, Q_J)$,
there are four regions of interest, and the behaviour of $\dot
\phi$ is qualitatively different in each region:
\begin{description}
\item[Region I:]
$Q_R>0$,
$Q_{J,\pi}^{\mathrm{crit}}<Q_J<Q_{J,0}^{\mathrm{crit}}$. The
function $V_\phi(\theta,Q_J,Q_R)$ goes from $-\infty$ at
$\theta=0$ to $\infty$ at $\theta=\pi$. From the monotonicity
property it follows that this function has a single zero.
\item[Region II:]
$Q_{J,\pi}^{\mathrm{crit}}<Q_J$,
$Q_{J,0}^{\mathrm{crit}}<Q_J$. The function
$V_\phi(\theta,Q_J,Q_R)$ goes from $\infty$ at $\theta=0$ to
$\infty$ at $\theta=\pi$. From the monotonicity property it is
possible to show that this function is positive definite and
has a single minimum.
\item[Region III:]
$Q_R<0$,
$Q_{J,0}^{\mathrm{crit}}<Q_J<Q_{J,\pi}^{\mathrm{crit}}$. The
function $V_\phi(\theta,Q_J,Q_R)$ goes from $\infty$ at
$\theta=0$ to $-\infty$ at $\theta=\pi$. From the monotonicity
property it follows that this function has a single zero.
\item[Region IV:]
$Q_J<Q_{J,\pi}^{\mathrm{crit}}$, $Q_J <
Q_{J,0}^{\mathrm{crit}}$. The function
$V_\phi(\theta,Q_J,Q_R)$ goes from $-\infty$ at $\theta=0$ to
$-\infty$ at $\theta=\pi$. From the monotonicity property it
is possible to show that this function is negative definite
and has a single maximum.
\end{description}
Similar results can be obtained for the velocities $\dot\psi$,
but these are omitted here.

\begin{figure}
\begin{center}
\includegraphics[scale=0.42]{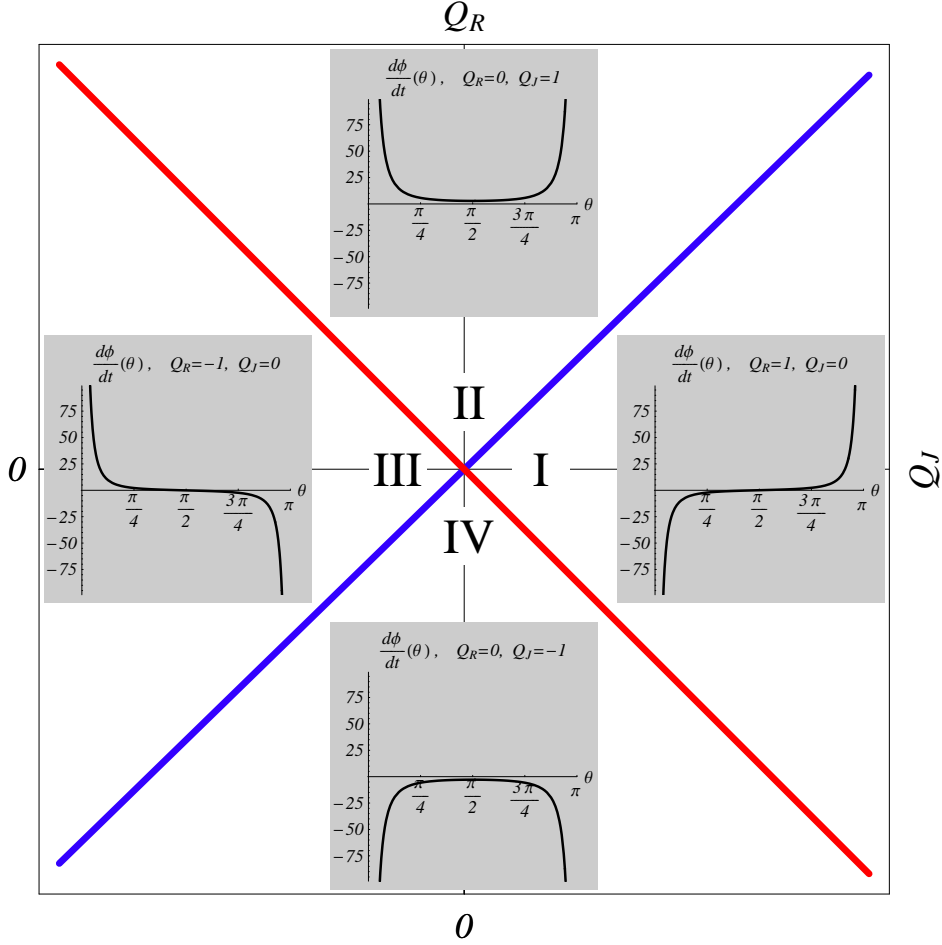}
\caption{(Colour online). The four critical regions defined by
the critical lines $Q_J = Q_{J,\pi}^{\mathrm{crit}}$ (red, top
left to bottom right) and  $Q_J = Q_{J,0}^{\mathrm{crit}}$
(blue, bottom left to top right). In each region, the graph of
$\dot \phi$ as function of $\theta$ is shown, for selected
values $Q_R, Q_J$ and parameters $a=0.05, I_3 = 2/5$ and
$I_1=(1-5\,a/2)I_3$.} \label{fig:4Regions}
\end{center}
\end{figure}

In figure~\ref{fig:4Regions} we show the four regions,
separated by the two critical lines $Q_J =
Q_{J,\pi}^{\mathrm{crit}}$ (solid red line from top left to
bottom right) and $Q_J = Q_{J,0}^{\mathrm{crit}}$ (dashed blue
line, from bottom left to top right). Typical plots of the
function $V_\phi(\theta,Q_J,Q_R)$ versus $\theta$ are inserted
in each region. The asymptotic behaviours are evident.

The critical Jellett quantity $Q_{J,\pi}^{\mathrm{crit}}$
plays a key role in determining the visible precession angle
$\Delta\phi\,(\mathrm{mod}\,2\pi)$ in the interesting case
$\theta_X\approx\pi$. The main contribution to the precession
angle comes from the turning point $\theta=\theta_X$ and from
(\ref{eq:phidotCJQR th=pi}) we see that the sign of this
contribution depends on the sign of $Q_J -
Q_{J,\pi}^{\mathrm{crit}}$. For example, if an initial
condition with $Q_J-Q_{J,\pi}^{\mathrm{crit}} \gtrapprox 0$
(Regions I or II) has a precession angle $\Delta\phi =
\alpha_0 \,(\mathrm{mod}\,2\pi)$, then a slightly different
initial condition with $Q_J-Q_{J,\pi}^{\mathrm{crit}}
\lessapprox 0$ (Regions IV or III) will have a precession
angle $\Delta\phi = -\alpha_0\,(\mathrm{mod}\,2\pi)$; the
corresponding motion will appear to be reversed.

The critical Jellett quantity $Q_{J,0}^{\mathrm{crit}}$
determines the full precession angle $\Delta\phi$ when
$\theta_N \approx 0$. The main contribution to $\Delta \phi$
comes from the turning point $\theta_N$, and is given by rapid
changes of $\phi$ in jumps of approximately $\pm\pi$, the sign
of these jumps depending on the sign of
$Q_J-Q_{J,0}^{\mathrm{crit}}$. In this way, an initial
condition in Region I or IV will give rise to a full-angle
precession $\Delta\phi<0$ whereas, for initial conditions in
Region II or III, $\Delta\phi>0$.
It is worth mentioning that this critical quantity is related
to the energy of the system since it appears in the Laurent
expansion of the function $F(\theta)$ near $\theta=0$. See
\cite{GrayNickel00}, where this critical quantity was
identified in terms of the centrifugal barrier.

\subsection{Quantitative estimate of applicability of
            criticality criteria}
\label{sec:range_X}

Let us consider the asymptotic region $\theta\approx\pi$. For
the above asymptotic analysis to be of practical importance,
the maximum rocking angle $\theta_X$ must be close to $\pi$.
Only then will the asymptotic Laurent expansion
(\ref{eq:phidotCJQR th=pi}) determine, to a good
approximation, the value of $\dot \phi$ at $\theta=\theta_X$.
In particular, we will observe a dramatic difference in $\dot
\phi$ at $\theta=\theta_X$ and in the precession angle when
considering two nearby points, one in Region I and one in
Region IV.

To quantify how close should $\theta_X$ be to $\pi$,
\textit{necessary} conditions are: (i) In Regions I and III,
$\theta_{\mathrm{z}}$, the zero of $\dot\phi(\theta)$, must be
less than $\theta_X$; (ii) In Regions II and IV,
$\theta_{\mathrm{e}}$, the extremum of $\dot\phi(\theta)$,
must be less than $\theta_X$. In each case, there is a
relation between $\theta_X, Q_R, Q_J$ and the parameters $a,
I_1, I_3.$

For Regions I and III, this condition has a simple analytical
formulation:
\[
\fl -1 \leq \cos\theta_X \leq \cos\theta_{\mathrm{z}} \equiv
\frac{a\beta\IC + \sqrt
{\IA(\IC-\beta)[\beta(\IC-\IA)(\IC+1)-a^2\beta\IC+\IA\IC(\IC+1)]}}
{\beta(\IC-\IA)+\IA\IC}
\]
where $\beta=\IC-(Q_J/Q_R)^2$. Realistic values of parameters
and ratio $Q_J/Q_R$ allow any value of $\theta_X$ in the
interval $(0,\pi)$.

It is noteworthy that, near the critical line $Q_J =
Q_{J,\pi}^{\mathrm{crit}}$, the necessary condition is
satisfied if and only if $\theta_X \approx \pi$. Letting
$Q_J/Q_{J,\pi}^{\mathrm{crit}} = 1 - \delta$ we get
 \begin{equation}
 \label{eq:range_X}
\pi \geq \theta_X \geq \pi
 - \left(\sqrt{\frac{2(1+a)[(1+a)^2+\IA]\IC}
                    {\IA(\IC+1+a)} } \,\right) {\delta}^{1/2}
+ O (\delta^{3/2}).
 \end{equation}

%
%


\section{Recession of the Asymmetric Body ($\bIA\ne\bIB$)}
\label{sec:Recession}

In this section we give a precise description of recession, or
reversal of precession, of the \RnR. The definition is based
on observational evidence: for initial conditions close to
pure rocking motion and such that the local maxima (turning
points) of the tilting angle $\theta$ are in the range
$(\thicksim 3/4 \pi,\pi)$, the rates of change $\dot
\phi(t_j)$ and $\dot \psi(t_j)$ at times
$t_j,\,j=1,...,\infty$ where the angle $\theta(t_j)$ is a
(local) maximum $\theta_X(t_j)$, depend on the time $t_j$,
contrary to the case of the symmetric body. The functions
$\dot \phi(t_j)$ and $\dot \psi(t_j)$ have a quasi-periodic
behaviour, undergoing changes of sign that translates
observationally to alternating reversals of the visible
precession angles $\Delta\phi(t_j)\,(\mathrm{mod}\,2\pi)$ and
$\Delta\psi(t_j)\,(\mathrm{mod}\,2\pi)$, where
\[
\Delta\phi(t_j)=\int_{t_{j-2}}^{t_j}\dot\phi(t)\,\rmd t\,,
\qquad \Delta\psi(t_j)=\int_{t_{j-2}}^{t_j}\dot\psi(t)\,\rmd
t\,.
\]
Note that the integration is from $t_{j-2}$ to $t_j$, which
accounts for a full period of motion.
In the dynamical region of interest,
$\theta_X(t_j)\in(\thicksim 3/4\pi,\pi)$, the critical
quantities defined in \S\ref{sec:criticality} allow us to
understand the behaviour qualitatively, and to predict the
occurrence of reversals.

The key observation from numerical simulations is that, in the
asymmetric case, the Jellett and Routh quantities,
(\ref{eq:jellett}) and (\ref{eq:routh}), cease to be
conserved, but {\em oscillate} about mean values. We thus
\emph{define} the Jellett and Routh quantities $Q_J$ and $Q_R$
to be
\begin{equation}
Q_J(t) = \IA s\omega_2^\prime+\IC f\omega_3^\prime\,, \qquad
Q_R(t) = \frac{\omega_3^{\prime}}{\rho} \,.
\label{eq:JelletandRouth_Qtys}
\end{equation}
We have observed that these quantities oscillate about
time-averaged values with a period that is generally longer
than the period of the rocking motion, and that depends on the
amplitude of the motion. We remark that the motion in the
$(Q_R,Q_J)$-plane is bounded. We will perform a numerical
study of this behaviour in connection with reversals at the
end of next section. The analytical study of this will be the
subject of forthcoming work.

The analysis in \S\ref{sec:criticality} regarding the
asymptotic behaviour of $\dot\phi$ near the turning points,
remains valid if we consider $Q_J(t)$ and $Q_R(t)$ to be
functions of time. In particular, as long as the point
$(Q_R(t), Q_J(t))$ remains within one of the Regions I to IV,
we can safely conclude that there is no reversal of the
system, because the sign of $\dot\phi$ at the turning points
cannot possibly change.
Reversal is due to crossing of the system from one region to
an adjacent one. In order to observe reversal, we need to
initialize the system sufficiently close to the boundary of a
region in such a way that, during the evolution of the motion,
the system crosses the boundary. We call this a
\textit{critical crossing}. Due to the oscillating nature of
$Q_J(t)$ and $Q_R(t)$ evidenced in numerical simulations, if
this critical crossing happens then the system will eventually
cross back to the original region and will continue crossing
periodically back and forth between the two regions, in a
bounded motion within the space $(Q_R, Q_J)$.

Corresponding to the critical crossings of the two types of
critical quantities --- $Q_{J,0}^{\mathrm{crit}}$ defined at
the turning point near $\theta=0$ and
$Q_{J,\pi}^{\mathrm{crit}}$ defined at that near $\theta=\pi$
--- there are two types of reversal. On the one hand, the {\em
full angle}, $\phi(t)$, has reversals that are related to the
critical crossings of $Q_{J,0}^{\mathrm{crit}}$. This is due
to the fact that, for motion close to pure rocking, the main
change of $\phi(t)$ from $t=t_{j-1}$ to $t=t_j$ is typically a
jump of magnitude about $\pi$ when $\theta$ passes the turning
point $\theta_N$. The sign of this jump depends on which
critical region the system is in, and will therefore change
when reversal occurs. Critical crossings from Region I to
Region II or from Region III to Region IV correspond to this
type of reversal.

On the other hand, the {\em visible precession angle}, $\Delta
\phi(t_j)\,(\mathrm{mod}\,2\pi)$, is due mainly to the change
of $\phi(t)$ near the turning point $\theta_X$. The sign of
this change depends exclusively on the criticality
$Q_{J,\pi}^{\mathrm{crit}}$. Critical crossings from Region IV
to Region I or from Region II to Region III determine this
type of reversal. This reversal corresponds to the recession
evident in real experiments.
A numerical study of the two types of reversal will be
presented in \S\ref{sec:numerics} below.

\section{Asymmetric Body ($\bIA\ne\bIB$): The Dynamical Equations}
\label{sec:AsymDynamics}

We now derive the equations for the asymmetric case
$\IA\ne\IB$. Since the intermediate frame $(\bi{i^\prime,
j^\prime, k^\prime)}$ is no longer a principal frame, it is
convenient to use a body frame $(\bi{i, j, k)}$ aligned in the
direction of the principal axes. The angular velocity and
angular momentum are then
\[
{\bomega} = \omega_1\bi{i} + \omega_2\bi{j} + \omega_3\bi{k}
\qquad \bi{L} = \IA\omega_1\bi{i} + \IB\omega_2\bi{j} +
\IC\omega_3\bi{k} \,.
\]
The momentum equations of motion in the body frame are
\begin{eqnarray}
\dot v_1 + \omega_2 v_3 - \omega_3 v_2 &=& F_1 \nonumber \\
\dot v_2 + \omega_3 v_1 - \omega_1 v_3 &=& F_2
\label{eq:Atrans} \\
\dot v_3 + \omega_1 v_2 - \omega_2 v_1 &=& F_3 \nonumber
\end{eqnarray}
and the angular momentum equations are
\begin{eqnarray}
\IA\dot\omega_1 + (\IC - \IB)\omega_2\omega_3 &=& G_1
\nonumber \\
\IB\dot\omega_2 + (\IA - \IC)\omega_3\omega_1 &=& G_2
\label{eq:Arotat} \\
\IC\dot\omega_3 + (\IB - \IA)\omega_1\omega_2 &=& G_3
\nonumber
\end{eqnarray}
We proceed as in \S\ref{sec:SymmDynamics}, using the
constraint (\ref{eq:constraint}) to express the velocity
$\bi{v}$ in terms of the rotation $\bomega$ and eliminating
the moment $\bi{G}$ by means of the momentum equations. The
result may be written
\begin{equation}
\bSigma\bthdot = {\bomega} \,, \qquad \bi{K}\bomdot
               = \bi{P}_{\bomega} \label{eq:thetaomega}
\end{equation}
where
$$\fl \bthdot = \left( \begin{matrix}
 { \dot\theta \cr \dot\phi \cr \dot\psi }
\end{matrix}  \right) \,, \qquad
\bomdot = \left( \begin{matrix}
 { \dot\omega_1 \cr \dot\omega_2 \cr \dot\omega_3 }
\end{matrix}  \right) \,,
$$
the matrices $\bSigma$ and $\bi{K}$ are
\begin{equation}
\fl \bSigma = \left[ \begin{matrix}
  { \chi   &  s\sigma  &  0  \cr
  -\sigma  &   s\chi   &  0  \cr
     0     &     c     &  1   }
\end{matrix}  \right]
\qquad \bi{K} = \left[ \matrix{
        \IA+f^2+s^2\chi^2 & -s^2\sigma\chi      & -fs\sigma \cr
         -s^2\sigma\chi   & \IB+f^2+s^2\sigma^2 & -fs\chi   \cr
            -fs\sigma     & -fs\chi             &  \IC+s^2  }
\right]\,,
\end{equation}
and the vector $\bi{P}_{\bomega}$ is
\[
\bi{P}_{\bomega} = \left(
\begin{array}{l}
       -(g+\omega_1^2+\omega_2^2)as\chi
       +(\IB-\IC-af)\omega_2\omega_3  \cr
\phantom{+}(g+\omega_1^2+\omega_2^2)as\sigma
       +(\IC-\IA+af)\omega_1 \omega_3 \cr
\phantom{+}(\IA-\IB)\omega_1\omega_2
       + as( -\chi\omega_1+\sigma\omega_2)\omega_3
\end{array}
\right)\,,
\]
with $\chi=\cos\psi$ and $\sigma=\sin\psi$. Note that neither
$\bi{K}$ nor $\bi{\bi{P}_{\bomega}}$ depends explicitly on
$\phi$. Thus $\phi$ is an ignorable coordinate in the system
(\ref{eq:thetaomega}).

\subsection{Special solutions}
\label{sec:spec-sol-asym}

We consider pure rocking motion with $\phi$ and $\psi$
constant. Then $\bomega=(\chi\dot\theta, -\sigma\dot\theta,
0)$. System (\ref{eq:thetaomega}) implies
\[
(\IA-\IB)\sigma\chi\dot\theta^2 = 0
\]
so that a nontrivial solution requires $\sigma\chi=0$. That
is, the rocking motion must be about one of the principal
axes. For $\sigma=0$ the system reduces to
(\ref{eq:purerock}), pure rocking about the $\bi{i}$-axis. For
$\chi=0$ we get the corresponding equation with $\IB$
replacing $\IA$ and pure rocking about the $\bi{j}$-axis. From
(\ref{eq:smallamprock}), the ratio of the small amplitude
oscillations about these principal axes is
\[
\frac{\nu_1}{\nu_2} = \sqrt{\frac{\IB+1}{\IA+1}} \,.
\]

In general, there are no periodic solutions corresponding to
the pure rolling motion found in the symmetric case.
However, if $\theta$ remains zero, we may have spinning about
the $\bi{k}$-axis, with $\bi{k}$ vertical. Then $\bomega=(0,
0, \dot\psi)$. The equations (\ref{eq:thetaomega}) reduce to
$\IC\dot\omega_3=0$, confirming that the spin rate is an
arbitrary constant.

\subsection{Nonholonomic constraints}
\label{sec:nonholo}

The \RnR\ is subject to three constraints, one holonomic and
two nonholonomic. The body must remain in contact with the
underlying surface, and the point of contact must be
momentarily stationary to ensure rolling contact. We can
embrace the three constraints in the single equation
(\ref{eq:constraint}), i.e, $\bi{v} = \bi{ \bomega\cross r}$.
We will now express this in terms of the space frame. The
velocities in the body and space frames, $\bi{v}$ and $\bi{V}$
respectively, are related by $\bi{v = \bUpsilon^{\rm T} V}$
or, explicitly,
\begin{equation}
\left( \begin{matrix} { v_1 \cr v_2 \cr v_3 }
\end{matrix}  \right)
= \left[ \begin{matrix} { \cphi\chi-c\sphi\sigma &
-\cphi\sigma-c\sphi\chi &  s\sphi \cr \sphi\chi+c\cphi\sigma
& -\sphi\sigma+c\cphi\chi & -s\cphi \cr
       s\sigma          &             s\chi        &     c    }
\end{matrix}  \right]^{\rm T}
\left( \begin{matrix} { V_1 \cr V_2 \cr V_3 }
\end{matrix}  \right)
\end{equation}
where $\cphi=\cos\phi$ and $\sphi=\sin\phi$. The matrix
$\bUpsilon$ is the product of three rotations, and is derived
in many standard texts in mechanics; see, for example,
\cite{HandFinch98,
      LandauLifshitz76,
      MontaldiRatiu05,
      SyngeGriffith59,
      Whittaker37}.
We write $\bi{\bomega}\cross\bi{r} = \bGamma\bi{\bomega}$ and
$\bi{\bomega} = \bSigma\bthdot$, where
\begin{equation}
\bGamma = \left[ \begin{matrix} {  0       &     f       &
-s\chi  \cr
   -f      &     0       &  s\sigma \cr
  s\chi    &  -s\sigma   &     0     }
\end{matrix}  \right] \,.
\label{eq:Gam}
\end{equation}
Now the constraints can be expressed in the form $\bi{V} =
\bUpsilon\bGamma\bSigma\,\bthdot$, relating the velocity in
the space frame to the time derivatives of the Euler angles.
More explicitly,
\begin{equation}
\left( \begin{matrix} { \dot X \cr \dot Y \cr \dot Z }
\end{matrix}  \right) =
\left[ \begin{matrix} {  h\sphi  &  -as\cphi   &  -s\cphi \cr
  -h\cphi  &  -as\sphi   &  -s\sphi \cr
    as     &     0       &     0     }
\end{matrix}  \right]
\left( \begin{matrix} { \dot\theta \cr \dot\phi \cr \dot\psi }
\end{matrix}  \right)
 \,.
\label{eq:XYZconstraints}
\end{equation}
It is clear from this form that the constraint on $\dot Z$ is
holonomic and may be integrated immediately to give $Z = 1-ac
= h$, the height of the mass centre in terms of the tilting
angle $\theta$. The constraints on $\dot X$ and $\dot Y$ are
nonholonomic.

\section{Lagrangian formulation}
\label{sec:Lagrange}

Systems with holonomic constraints can be solved by
elimination of redundant coordinates or by adding to the
Lagrangian a sum of the constraints weighted by Lagrange
multipliers. When the constraints are nonholonomic, this
procedure does not apply \cite{LandauLifshitz76, Whittaker37}.
We must resist the temptation to substitute
(\ref{eq:constraint}) into the Lagrangian and obtain a
Lagrangian that involves only the Euler angles and their
derivatives. Rather, we must embed the problem in a
configuration space of dimension $N+M$, where $N$ is the
number of degrees of freedom and $M$ the number of
nonholonomic constraints. There has been considerable
misunderstanding regarding nonholonomic constraints; see
\cite{Flannery05} for a review.  When the constraints are of
the form
\[
g_k({\bi{q},\bqdot,t}) \equiv
         \bi{A}_k({\bi{q},t}) \bqdot = 0  \,,
\]
that is, where they are linear in the velocities, we can write
the equations of motion in the form
\begin{equation}
\frac{\rmd}{\rmd t}\frac{\partial {\cal L}}{\partial\bqdot} -
\frac{\partial {\cal L}}{\partial\bi{q}} + \sum_k
\mu_k\frac{\partial g_k}{\partial\bqdot} = 0 \,.
\label{eqn:LagNonHolConstr}
\end{equation}
where $\mu_k$ are Lagrange multipliers that can be determined
using the constraints. In the present case, the configuration
space has five dimensions, with coordinates
$(\theta,\phi,\psi,X,Y)$, the holonomic constraint having been
used to eliminate $Z$. We may write the Lagrangian in terms of
these coordinates and their time derivatives:
\begin{eqnarray}
{\cal L} = \half \bigl\{
  & (\IA\chi^2+\IB\sigma^2+a^2s^2) \dot\theta^2
    + 2(\IA-\IB)s\chi\sigma\,\dot\theta\dot\phi \nonumber \\
  & + [(\IA\sigma^2+\IB\chi^2)s^2+\IC c^2]\dot\phi^2
    + 2\IC c\,\dot\phi\dot\psi + \IC\,\dot\psi^2  \\
  & + (\dot X^2 + \dot Y^2) \bigr\} - ga(1-c)  \nonumber \,.
\label{eq:Lagrangian}
\end{eqnarray}
Note that ${\cal L}$ does not depend on $\phi$.
From (\ref{eq:XYZconstraints}), the nonholonomic constraints
are
\begin{eqnarray}
g_1 \equiv &\dot X - ( \phantom{-}
 h\sphi\,\dot\theta  -as\cphi\,\dot\phi -s\cphi\,\dot\psi )
&= 0 \cr g_2 \equiv &\dot Y - ( -h\cphi\,\dot\theta
-as\sphi\,\dot\phi -s\sphi\,\dot\psi ) &= 0
\end{eqnarray}
Although $\phi$ occurs in these expressions, it is absent from
the combinations $\sum_k \mu_k{\partial g_k}/{\partial\bqdot}$
that occur in the equations.  This symmetry should imply the
existence of an invariant quantity in addition to the total
energy.

The Euler-Lagrange equations (\ref{eqn:LagNonHolConstr}) for
$X$ and $Y$ immediately yield
\[
  \mu_1 = - \ddot X \qquad \mu_2 = - \ddot Y \,.
\]
Using the constraints, we may now eliminate the multipliers
$\mu_1$ and $\mu_2$ from the remaining equations and obtain a
system of three equations for $\theta$, $\phi$ and $\psi$.
They may be written
\begin{equation}
\bi{M}{\bthddot} + {\bi{P}}_{\btheta}(\btheta,\bthdot) =
\mathbf{0} \label{eq:thtddot}
\end{equation}
where $\bthddot = (\ddot\theta,\ddot\phi,\ddot\psi)^{\rm T}$
and $\bi{M}$ is a symmetric matrix. The explicit expansion of
(\ref{eq:thtddot}) is given in the Appendix (\S A.3). Using
{\em Mathematica}, the system has been shown to be completely
equivalent to the system (\ref{eq:thetaomega}).


In general, we can write the Lagrangian in the form
\[
{\cal L} = {\cal L}_0(\theta,\bthdot) +
            \epsilon {\cal L}_1(\theta,\psi,\bthdot)
\]
where ${\cal L}_0$ is the Lagrangian for the integrable
symmetric system and $\epsilon\equiv(\IB-\IA)/\IA$ is the {\em
asymmetry parameter}. This provides a basis for a perturbation
analysis when $\epsilon$ is small, which will not be
undertaken here but will be the subject of future work.



\section{Numerical Experiments}
\label{sec:numerics}

The numerical integration of the equations is delicate, as
there is a singularity of the coordinate system when
$\theta=0$, and significant errors may result from this.
%
%
To be sure of reliable numerical results, we used a routine of
eighth-order accuracy, {\sc ode87}, coded by Vasiliy
Govorukhin ({\tt http://www.mathworks.com/matlabcentral/}),
which is a realization of the formulae of Prince and Dorman
\cite{PrinceDorman81}. With this method, invariants of the
motion remained constant to high accuracy. To further confirm
the robustness of the  numerics, we coded both sets of
equations, the system (\ref{eq:thetaomega}) in terms of
($\bthdot,\bomdot$) and the system (\ref{eq:thtddot}) in terms
of ($\bthdot,\bthddot$), and compared the results.
Furthermore, we verified the {\sc matlab} coding by an
independent coding in {\em Mathematica}. Finally, the results
presented below were checked for convergence by varying the
error tolerance. We can therefore be confident in the
reliability of the numerical results.

Unless otherwise stated, the numerical values of the
parameters are set as follows: gravity $g=9.87$, unit mass,
unit radius, centre of mass off-centering $a=0.05$, moments of
inertia $\IA=0.35$ and $\IC=0.4$. Some initial conditions will
not be varied in the various simulations; these are
$\theta_0=0.95\pi$, $\dot\theta_0=0$ and $\phi_0=0$.

\subsection{The consequence of asymmetry}

\begin{figure}
\begin{center}
\includegraphics[scale=0.38]{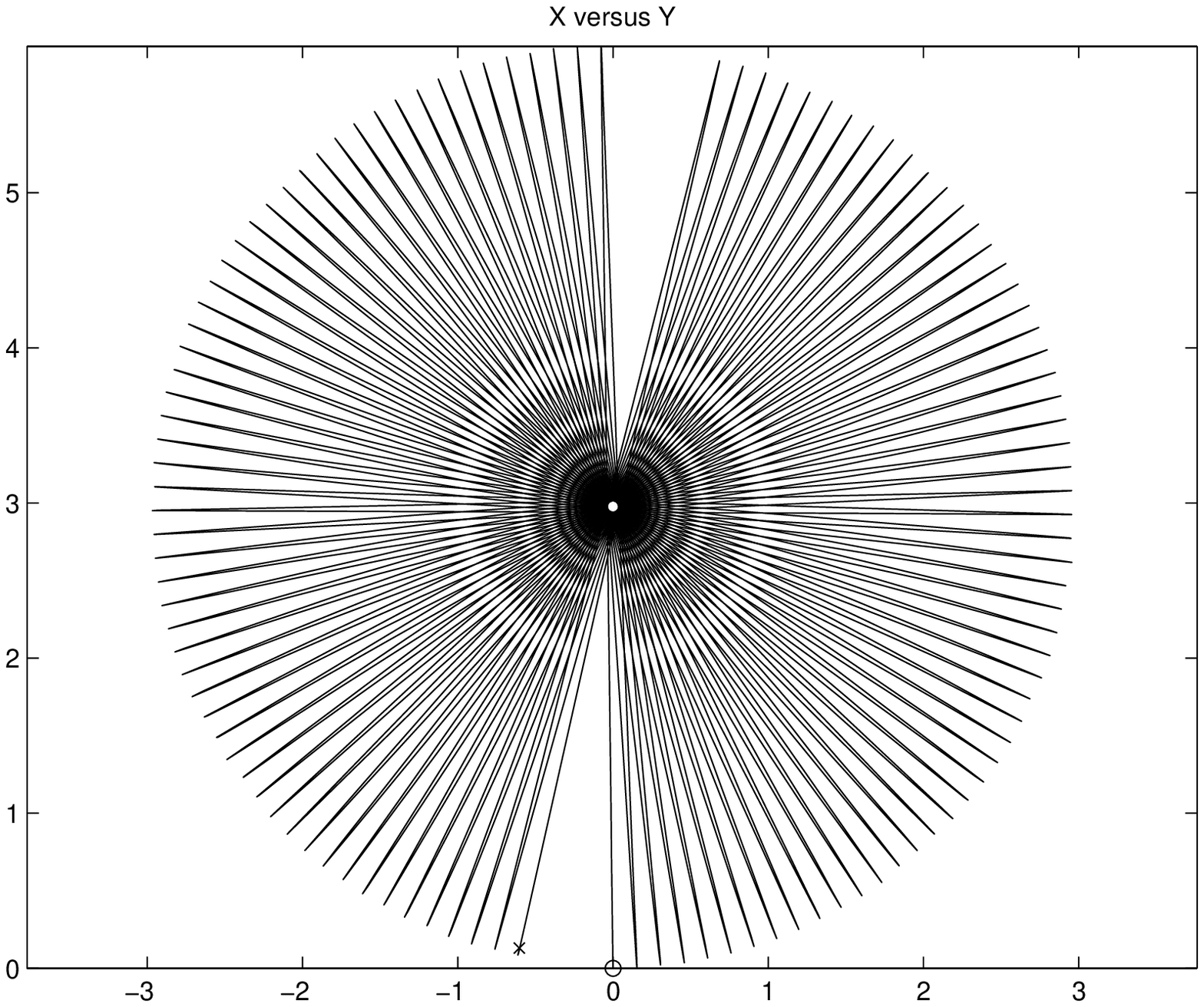}
\includegraphics[scale=0.38]{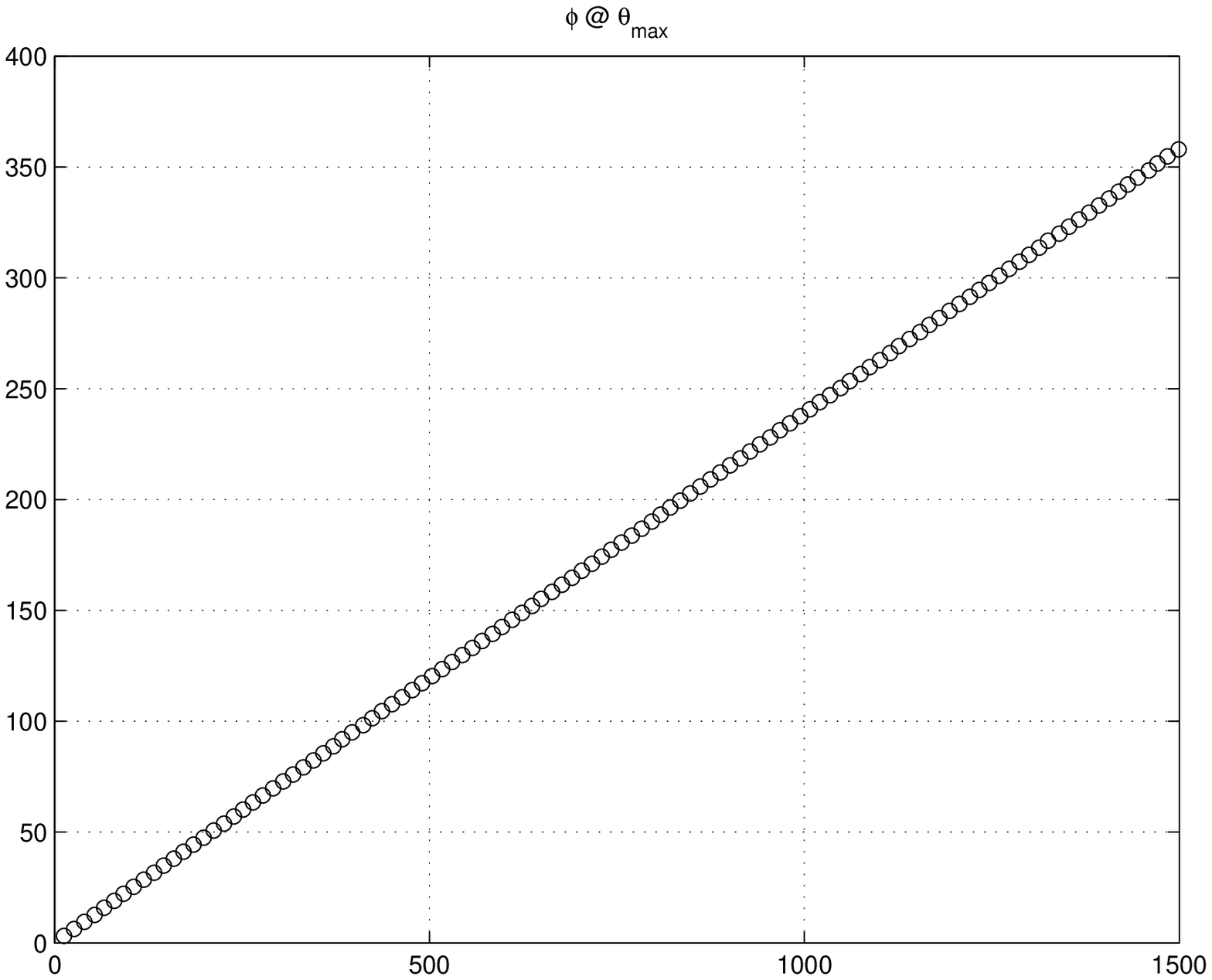}
\includegraphics[scale=0.38]{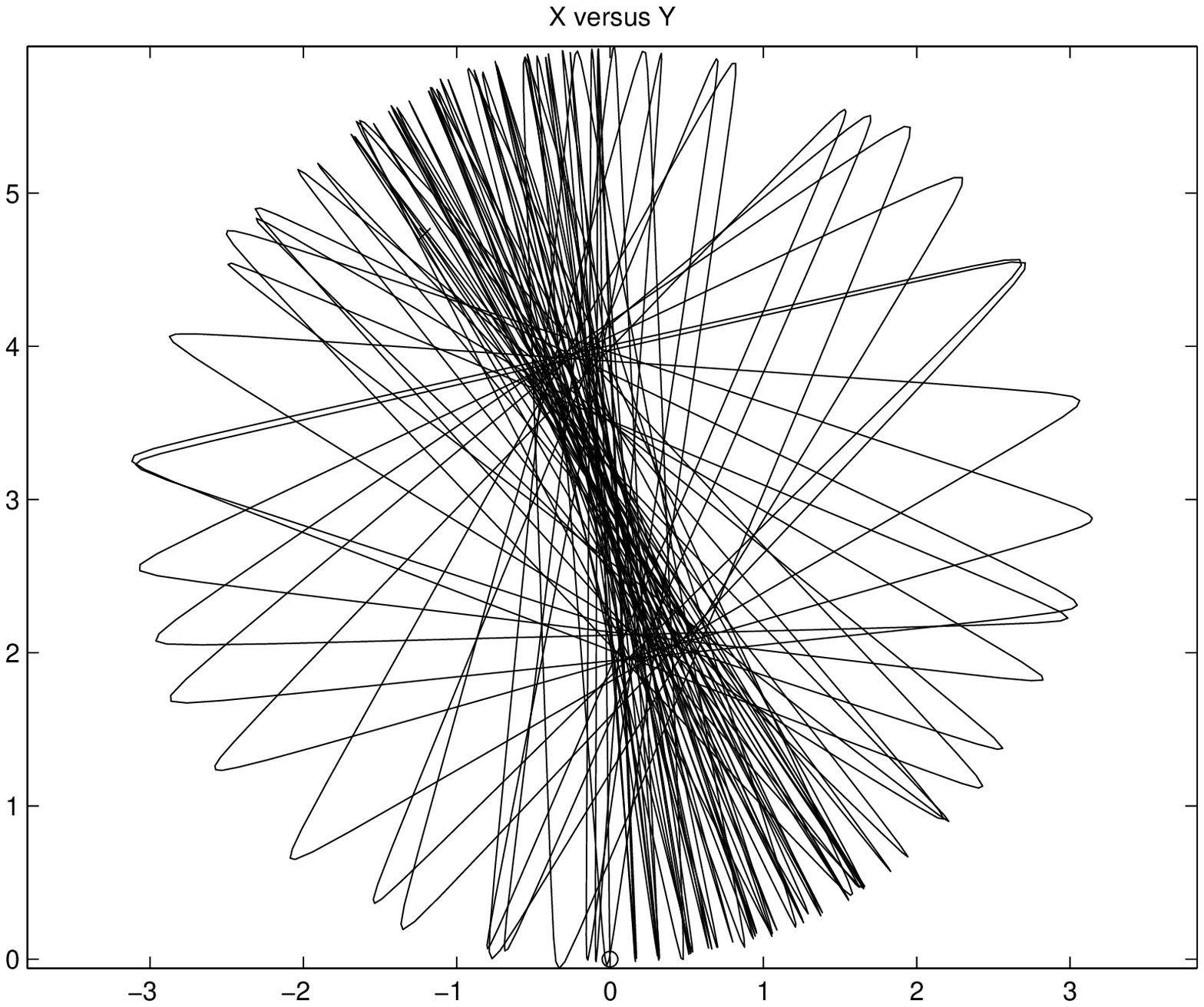}
\includegraphics[scale=0.38]{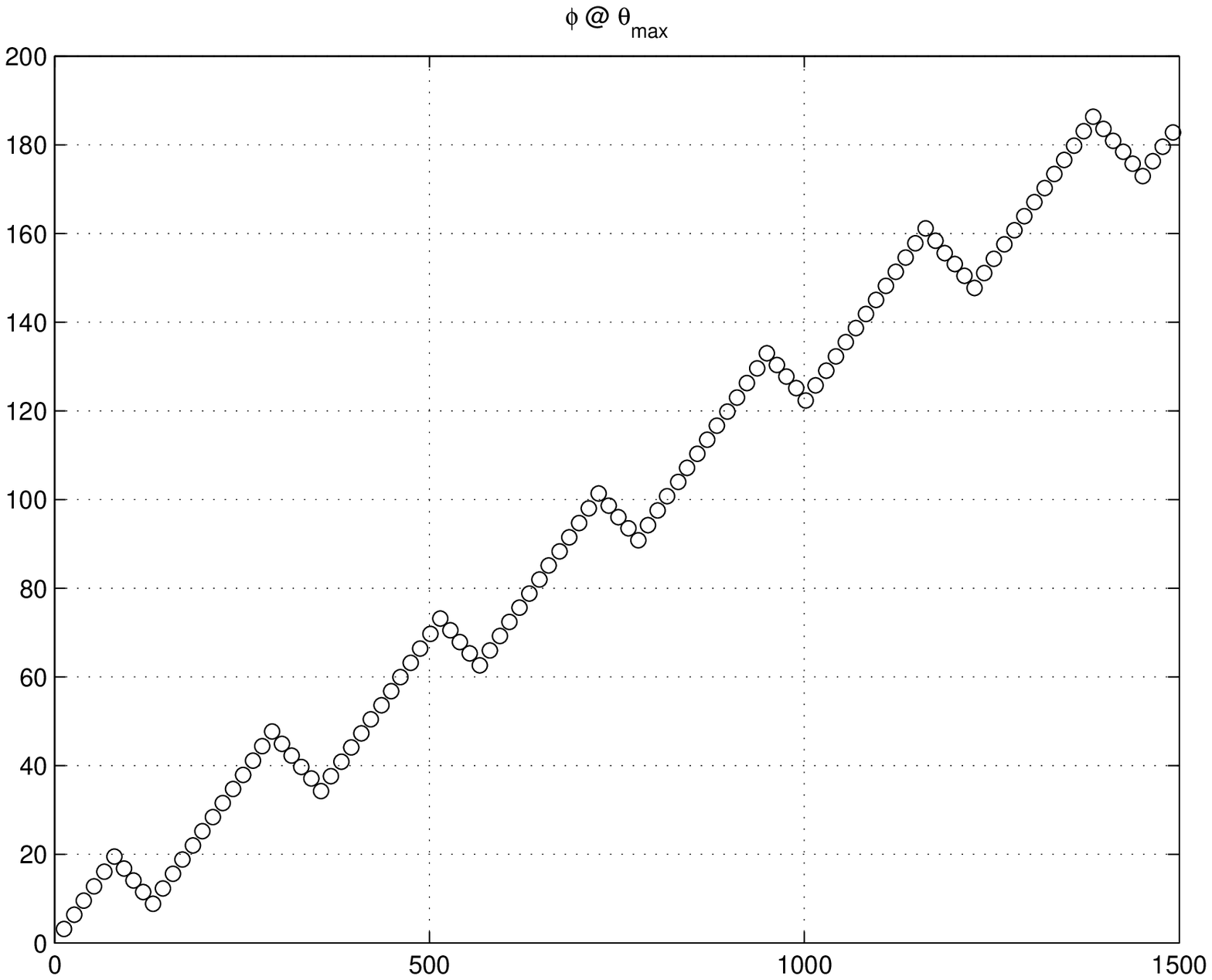}
\caption{Top left: Trajectory of the point of contact for
solution for symmetric case $\IA=\IB$. Top right: $\phi$
versus $\theta_{\rm max}$ for symmetric case. Bottom left and
right: corresponding solution for the asymmetric case
$\IB=1.001\IA$.} \label{fig:XYsymm}
\end{center}
\end{figure}

 We first compare the numerical solution of equations
(\ref{eq:thetaomega}) for the symmetric case $\IA=\IB$ and for
a case of slight asymmetry.  The solutions are for 1500 time
units and the initial conditions are, in each case,
$\theta_0=0.95\pi$, $\phi_0=\psi_0=0$, $\omega_{1,0}=0$,
$\omega_{2,0}=0.001$, $\omega_{3,0}=-0.001$.
Figure~\ref{fig:XYsymm} (top left panel) shows the trajectory
of the point of contact for the symmetric case $\IA=\IB$. The
azimuthal angle increases regularly and steadily for each
cycle of rocking motion. This is confirmed by
figure~\ref{fig:XYsymm}, top right panel, which shows $\phi$
at the points where $\theta$ reaches a maximum. For the
solution shown in the bottom panels of
figure~\ref{fig:XYsymm}, the only difference is a increase of
$0.1\%$ in the inertial moment about $\bi{j}$, so that
$\epsilon=(\IB-\IA)/\IA=10^{-3}$. The bottom left panel shows
the trajectory of the point of contact: the precession is no
longer uniform. The azimuthal angle alternately increases and
decreases (figure~\ref{fig:XYsymm} bottom right panel). We see
that there is recession, with a period much longer than that
of the rocking motion. Thus, a minute change in the mass
distribution of the body, that changes the inertial structure
slightly and breaks the symmetry $\IA=\IB$, has a dramatic
effect on the character of the motion.


\subsection{Stability of rocking motion}

We initiate the motion from a stationary state with
$\bomega(0)=0$ and $\theta(0)=0.95\pi$. Clearly, a symmetric
body started in this configuration would execute pure rocking
motion, passing repeatedly through the equilibrium position,
with $\phi$ and $\psi$ remaining constant (apart from jumps of
$\pi$ due to the coordinate singularity at $\theta=0$). For
the asymmetric body, the solution depends on the initial angle
$\psi(0)=\psi_0$. As before, we assume the asymmetry is
slight, with $\IB=1.001\IA$.

\begin{figure}
\begin{center}
\includegraphics[scale=0.84]{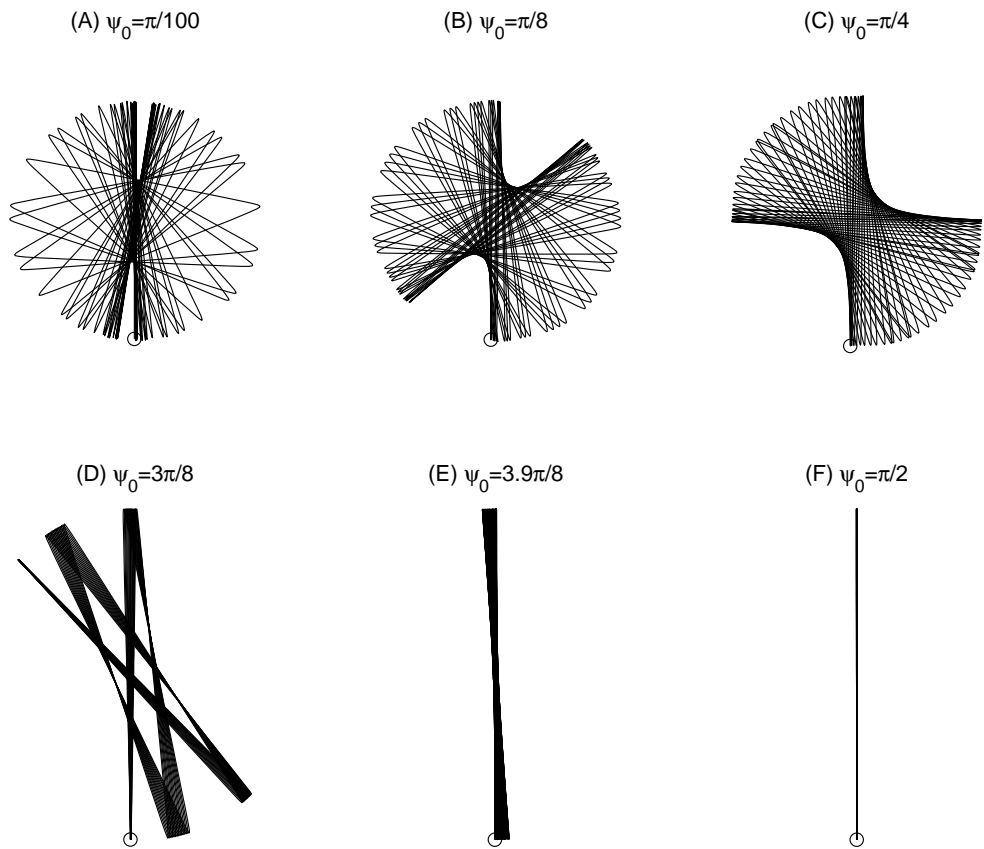}
\caption{Trajectory of the point of  contact in the $XY$-plane
for initial phase angle $\psi_0 \in \{\pi/100, \pi/8, \pi/4,
3\pi/8, 3.9\pi/8, \pi/2\}$. All integrations are for 1000 time
units. The small circles indicate the starting position in
each case.} \label{fig:Phi-span}
\end{center}
\end{figure}

The trajectory of the point of contact of the \RnR\ is shown
in figure~\ref{fig:Phi-span} for $\psi_0$ in the set
$\{\pi/100, \pi/8, \pi/4, 3\pi/8, 3.9\pi/8, \pi/2\}$. All
integrations are for 1000 time units. We see that the motion
precesses through an angle $\Phi=\phi_{\rm max}-\phi_{\rm
min}$ that depends sensitively on the initial phase $\psi_0$.
It appears that the relationship
\[
\Phi = \pi-2\psi_0  \qquad\mbox{for}\qquad \psi_0\in (0,\pi)
\]
is satisfied, at least approximately.

The cases $\psi_0=0$ and $\psi_0=\pi/2$ correspond to pure
rocking about the principal axes with moments of inertia $\IB$
and $\IA$ respectively. Motion close to pure rocking about the
$\IA$ axis is stable (figure~\ref{fig:Phi-span}(E)) while that
starting close to the $\IB$ axis changes dramatically,
precessing through almost $180^\circ$
(figure~\ref{fig:Phi-span}(A)). We recall the classical result
for free motions of a rigid body with $\IA<\IB<\IC$, where
rotation about the $\IB$ axis is unstable whereas rotations
about the $\IA$ and $\IC$ axes are stable.

In general we expect the trajectory to be dense in the domain
of angle $\Phi$ spanned by the solution. However, KAM theory
\cite{Berry78, Ott95} suggests that for exceptional initial
conditions the solution is periodic. The character of the
solution for $\psi_0=3\pi/8$ appears to be close to a periodic
solution (figure~\ref{fig:Phi-span}(D)). Searching in the
neighborhood of this solution, we found that when
$\psi_0=2.965\pi/8$ the trajectory becomes periodic,
repeatedly tracing out the same track, some fifteen times in
1000 seconds. Solutions of this nature, whose trajectories
span a set of measure zero, are a signature of integrability.



\subsection{Recession and Criticality}

\begin{figure}
\begin{center}
\includegraphics[scale=0.325]{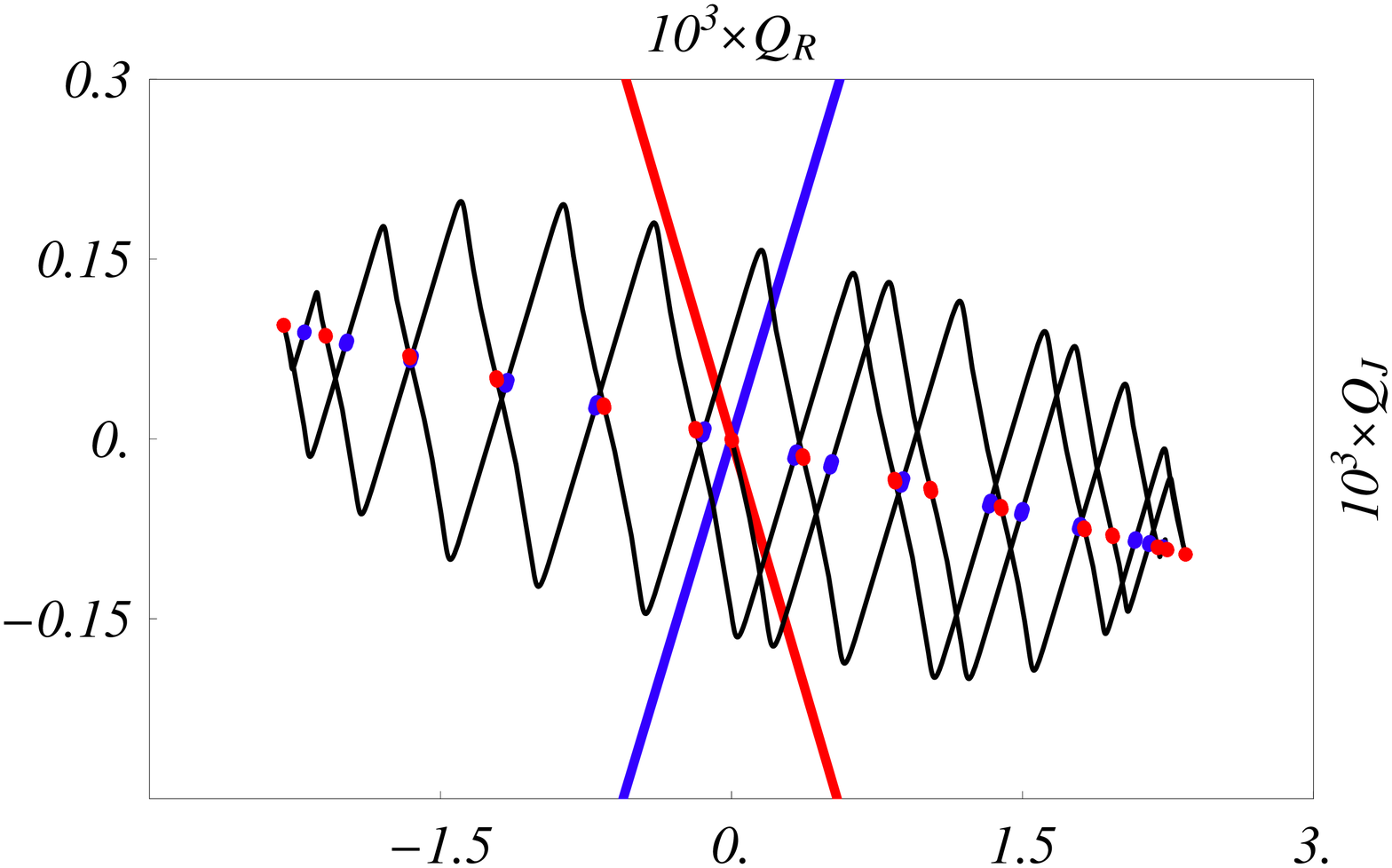}
\includegraphics[scale=0.325]{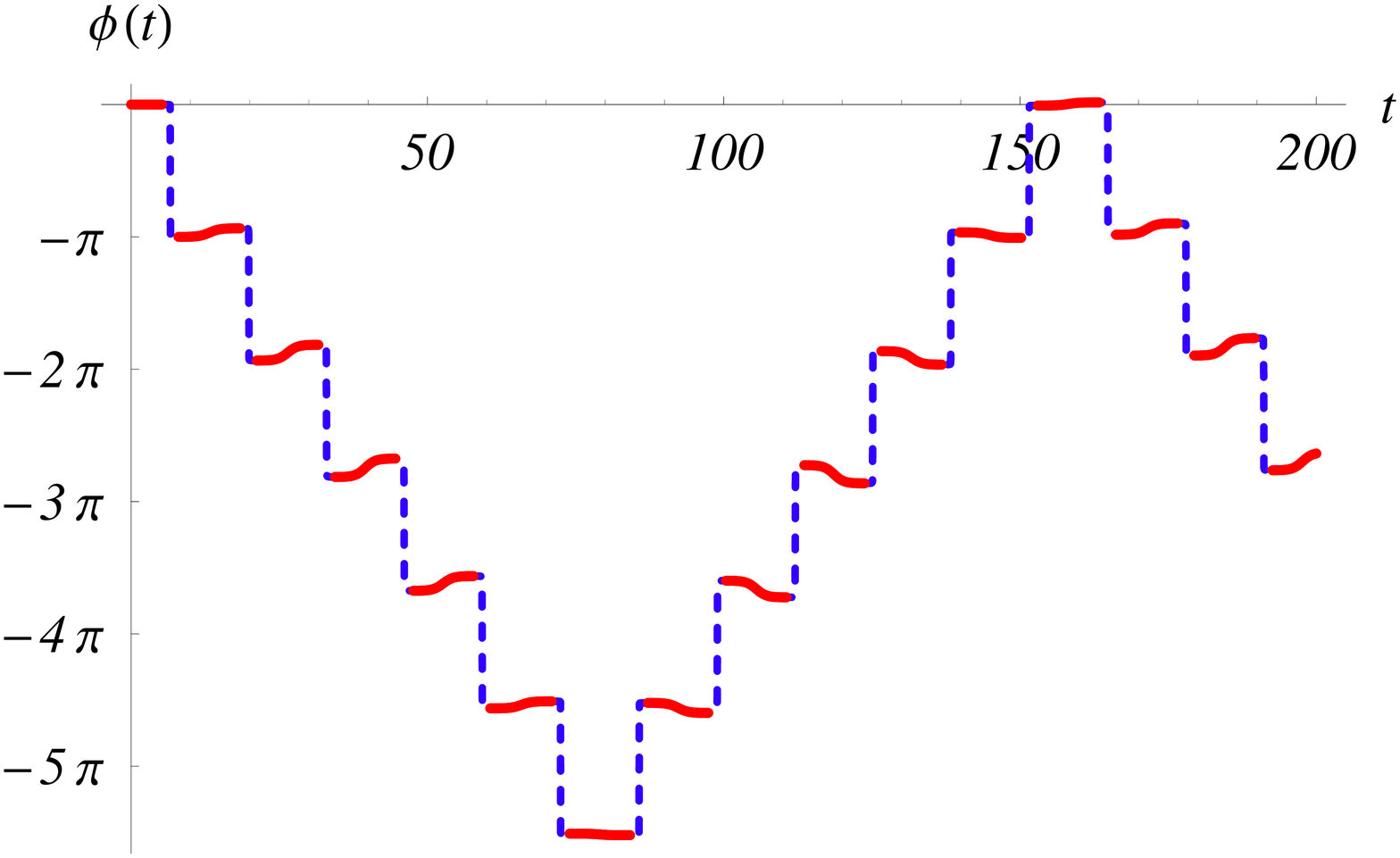}
\includegraphics[scale=0.325]{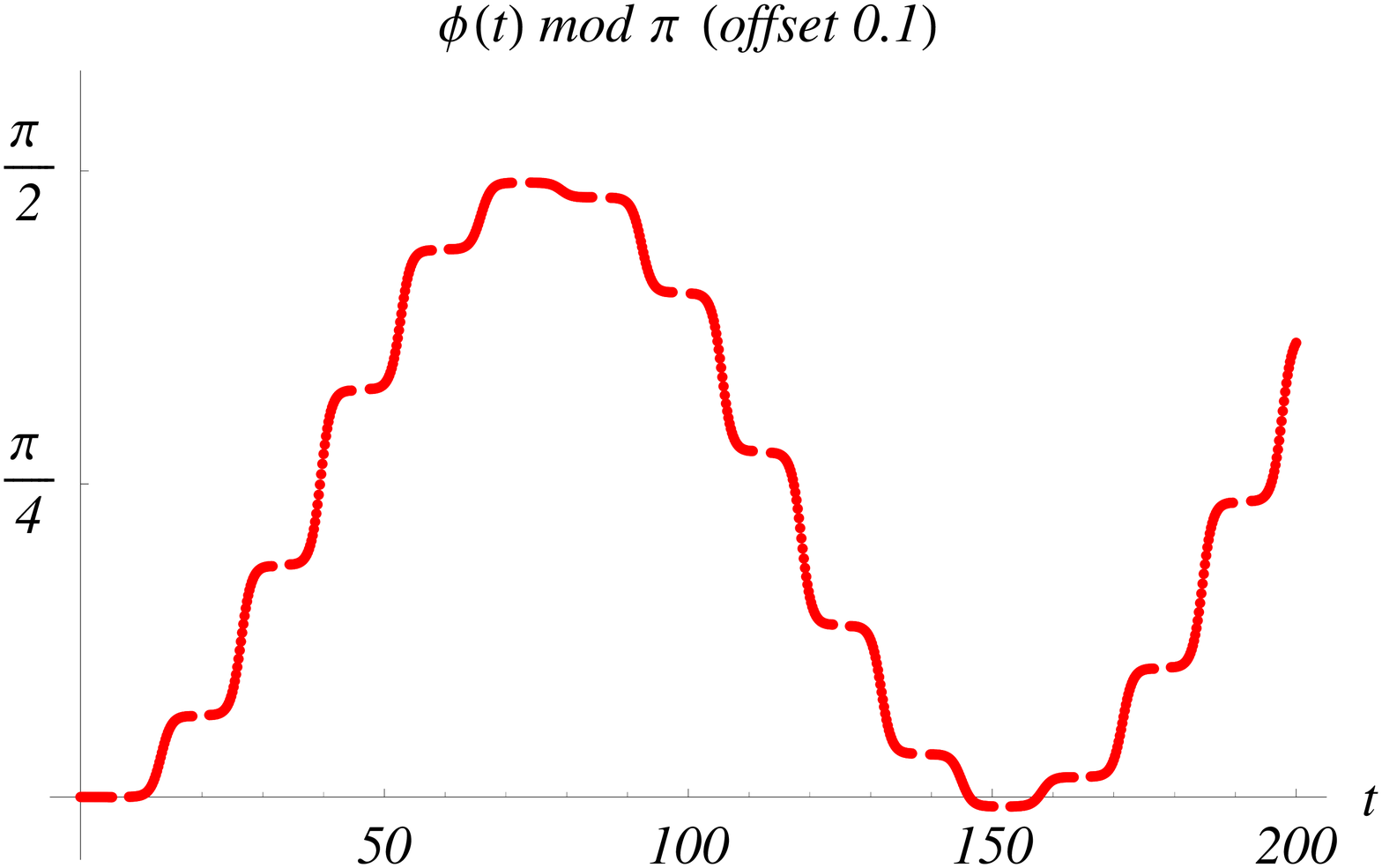}
\caption{(Colour online). Top frame: orbit in the
$(Q_R,Q_J)$-plane for 200 time units. Middle frame: Azimuth
angle $\phi(t)$. Bottom frame: Visible angle
$\phi(t)\,(\mathrm{mod}\,\pi)$ sampled when $\theta \geq
0.5\theta_0$. Initial conditions $Q_R=0, Q_J=0$. For full
details, see text.} \label{fig:reversals(QR,QJ)1}
\end{center}
\end{figure}

\begin{figure}
\begin{center}
\includegraphics[scale=0.325]{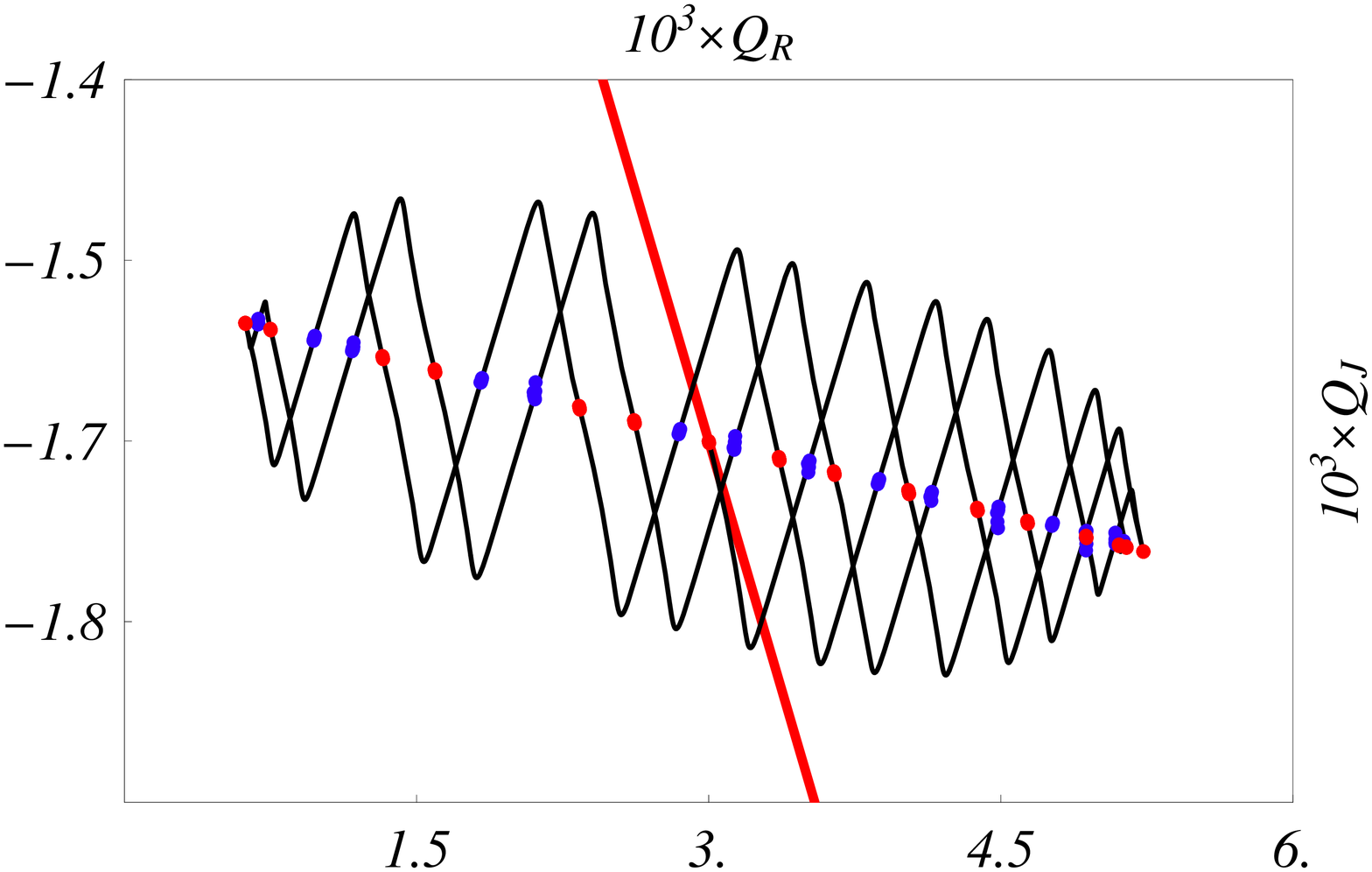}
\includegraphics[scale=0.325]{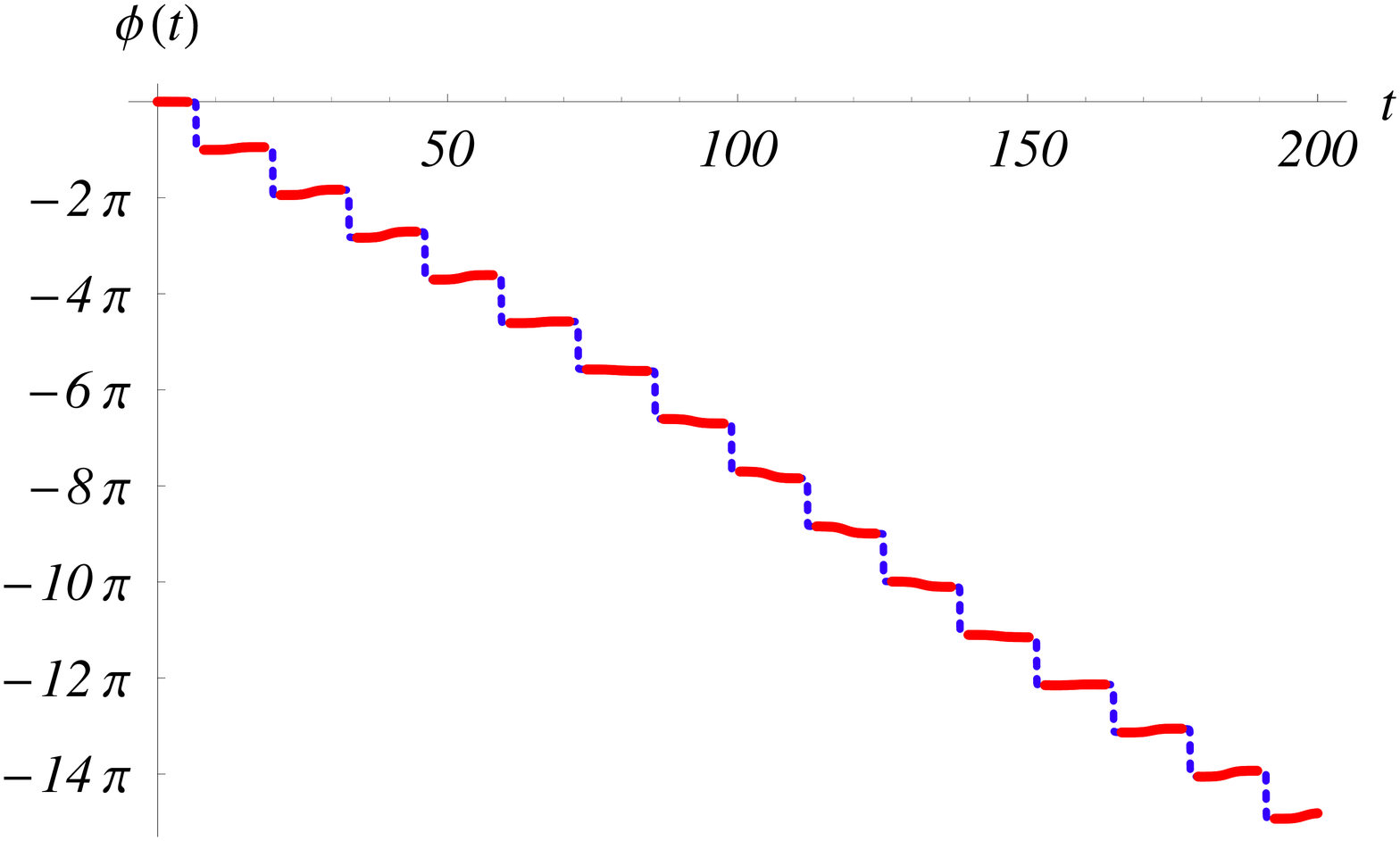}
\includegraphics[scale=0.325]{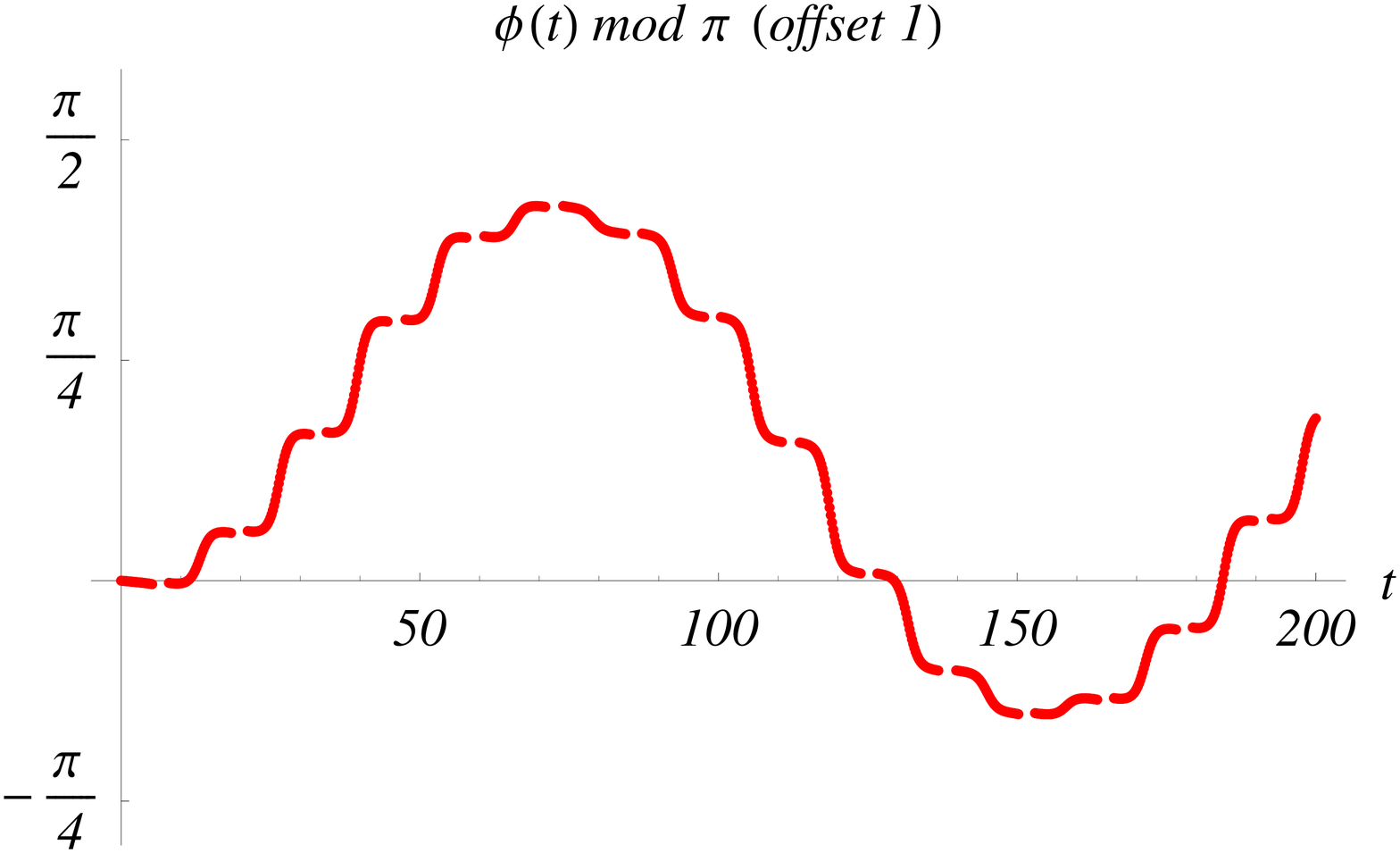}
\caption{(Colour online). Top frame: orbit in the
$(Q_R,Q_J)$-plane for 200 time units. Middle frame: Azimuth
angle $\phi(t)$. Bottom frame: Visible angle
$\phi(t)\,(\mathrm{mod}\,\pi)$ sampled when $\theta \geq
0.5\theta_0$. Initial conditions $Q_R=0.0030, Q_J=
Q_{J,\pi}^{\mathrm{crit}} \approx -0.0017$. For full details,
see text.} \label{fig:reversals(QR,QJ)2}
\end{center}
\end{figure}

\begin{figure}
\begin{center}
\includegraphics[scale=0.325]{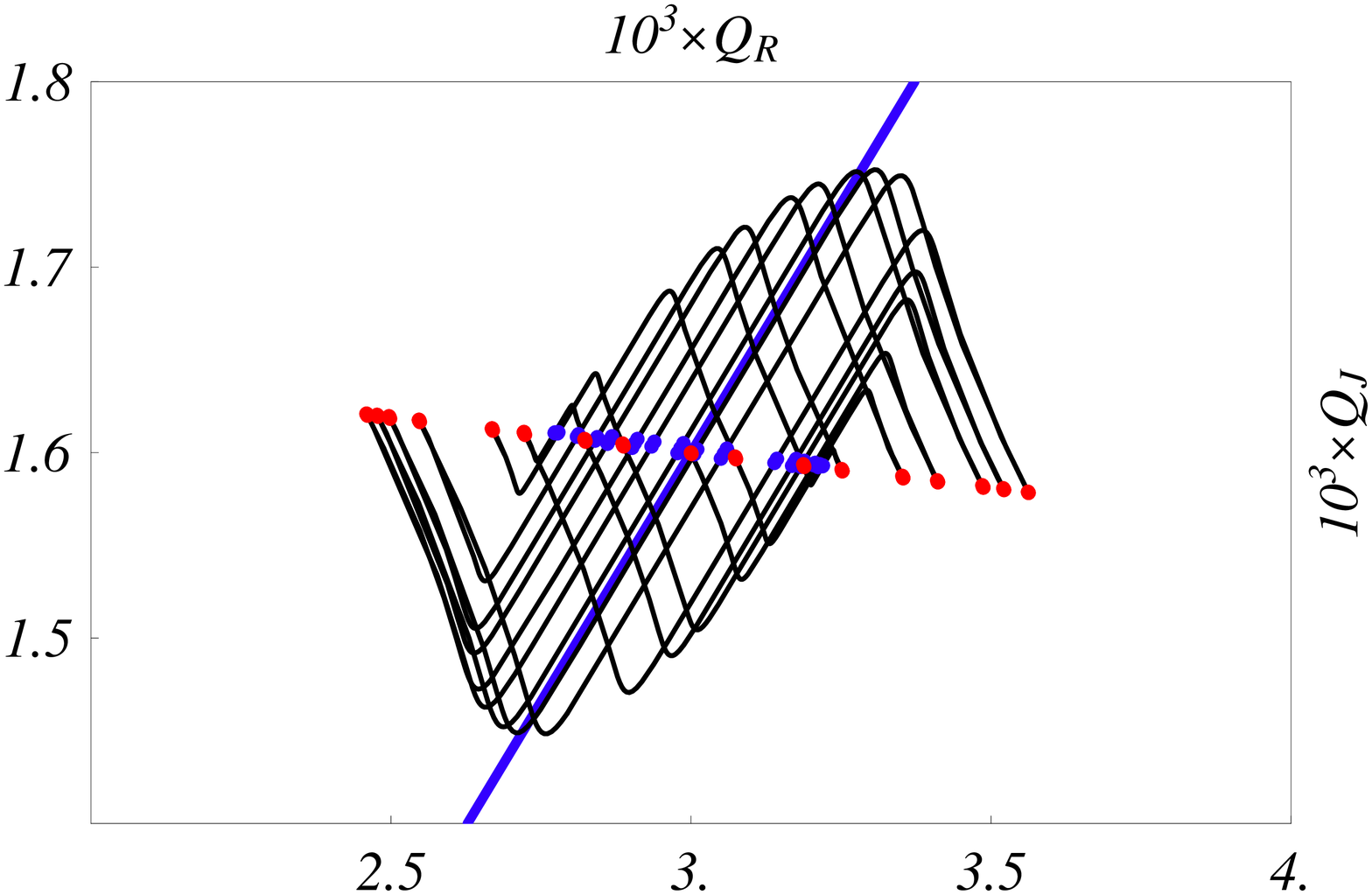}
\includegraphics[scale=0.325]{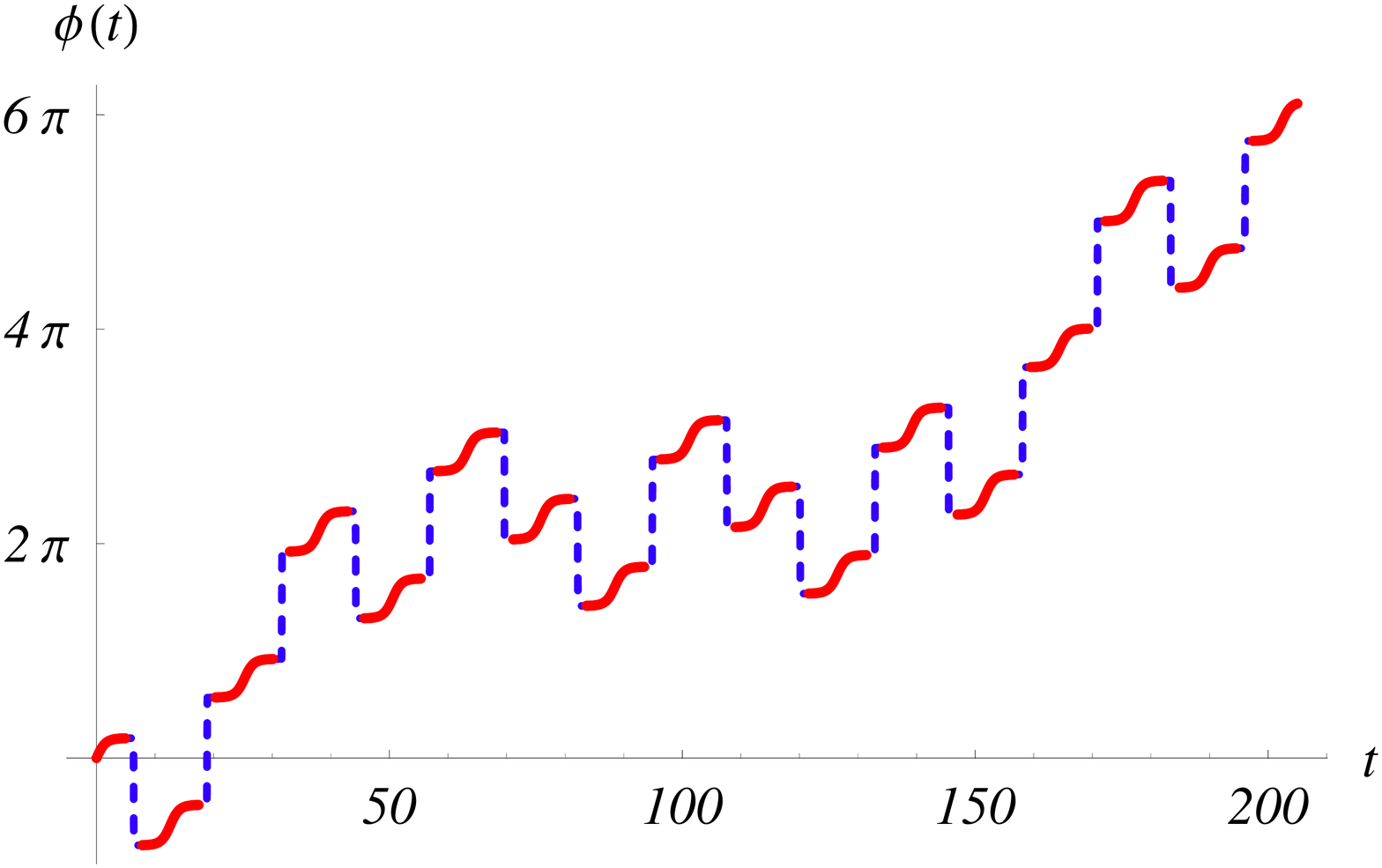}
\includegraphics[scale=0.325]{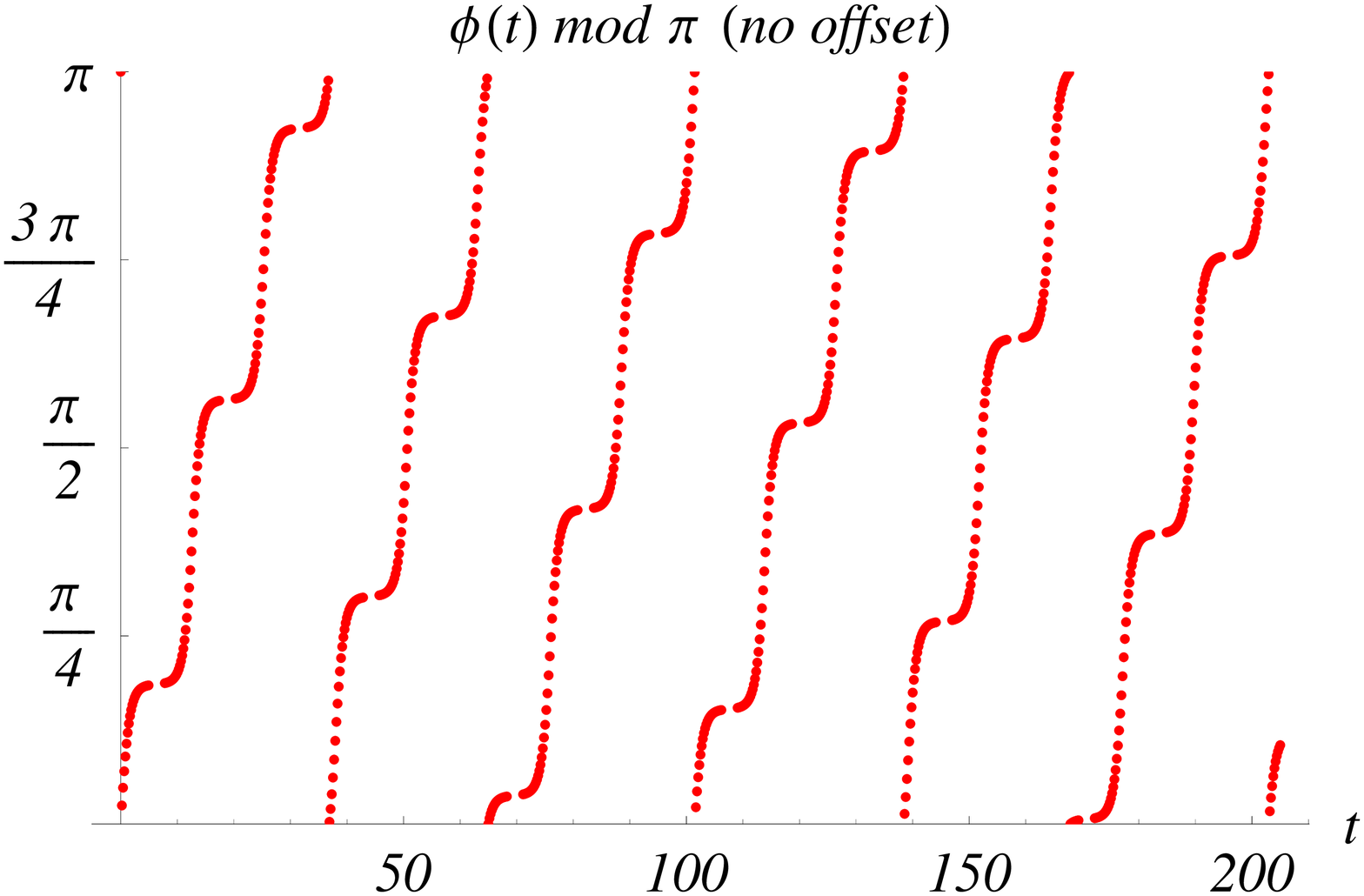}
\caption{(Colour online). Top frame: orbit in the
$(Q_R,Q_J)$-plane for 200 time units. Middle frame: Azimuth
angle $\phi(t)$. Bottom frame: Visible angle
$\phi(t)\,(\mathrm{mod}\,\pi)$ sampled when $\theta \geq
0.5\theta_0$. Initial conditions $Q_R=0.0030, Q_J=
Q_{J,0}^{\mathrm{crit}} \approx 0.0016$. For full details, see
text.} \label{fig:reversals(QR,QJ)3}
\end{center}
\end{figure}

We now present a numerical study of reversals based on the
theory of criticality described in \S\ref{sec:criticality}. In
all cases, the numerical experiments consist of releasing the
\RnR\ at an angle $\theta_0=0.95\pi$, with $\dot\theta_0=0$
and with an angle $\psi_0=\pi/4$ half-way between the body's
principal axes. System (\ref{eq:thetaomega}) is integrated
numerically for $200$ time units, using an adaptive {\em
Mathematica} code (stiffness-switching method) with $11$th
order accuracy. The results are insensitive to resolution
improvements. We monitor energy conservation point-wise and
confirm that the relative error is less than $10^{-10}$.
Routh's and Jellett's quantities are computed in
post-processing.

In figures~\ref{fig:reversals(QR,QJ)1},
\ref{fig:reversals(QR,QJ)2} and \ref{fig:reversals(QR,QJ)3}
all parameters are identical except for the initial
conditions. In all three cases, $\theta_0 = 0.95\pi$,
$\dot\theta_0=0$, $\phi_0=0$, $\psi_0=\pi/4$. Parameter values
are $a=0.05$, $g=9.87$, $\IC=2/5$, $\IA=(1-5 a/2)\IC$,
$\epsilon=(\IB-\IA)/\IA=10^{-3}$. The top frame in each case
shows the  orbit in the $(Q_R,Q_J)$-plane for 200 time units
(zigzagging bounded curve (black)). In all three cases, the
orbit starts at a point on one of the critical lines and
begins moving to the bottom right, subsequently alternating
between adjacent critical regions. The straight lines from top
left to bottom right (red) denote the critical line
$Q_J=Q_{J,\pi}^{\mathrm{crit}}$, useful for visible angle
reversal. The straight lines from bottom left to top right
(blue) denote the critical line $Q_J =
Q_{J,0}^{\mathrm{crit}}$, useful for full angle reversal. Dots
correspond to instances when $\theta(t)$ is near the turning
points: red dots denote $\theta\geq 0.99 \theta_0$ (near
turning point $\theta_X$) and blue dots denote
$\theta<0.005\theta_0$ (near turning point $\theta_N$). The
middle frame in each case shows the azimuth angle $\phi(t)$.
Solid lines (red) denote instances when $\theta \geq
0.5\theta_0$ and dashed lines (blue) denote instances when
$\theta < 0.5 \theta_0$. The bottom frames show the visible
(half) angle $\phi(t)\,(\mathrm{mod}\,\pi)$ sampled when
$\theta \geq 0.5\theta_0$.

In figure~\ref{fig:reversals(QR,QJ)1}, the initial velocities
are $\dot\phi_0=\dot\psi_0=0$. The motion remains close to the
centre $Q_R=Q_J=0$, i.e., close to pure rocking (black
zigzagging orbit in top frame). The system alternates between
Regions I and III, spending approximately five periods of
rocking motion in each region. Consequently, both full-angle
reversal (middle frame) and visible angle reversal (bottom
frame) can be observed. This case corresponds exactly to case
(C) in figure~\ref{fig:Phi-span}.

In figure~\ref{fig:reversals(QR,QJ)2}, the initial conditions
are right on the critical line
$Q_J=Q_{J,\pi}^{\mathrm{crit}}$: the initial velocities are
$\dot\phi_0=-0.002$ and $\dot \psi_0=0.002$. The orbit is very
similar in shape and size to the orbit in the previous case.
We observe two critical crossings between Region I and Region
IV, corresponding to visible angle reversals (bottom frame).
There is no full angle reversal (middle frame).

In figure~\ref{fig:reversals(QR,QJ)3}, the initial conditions
are right on the critical line $Q_J=Q_{J,0}^{\mathrm{crit}}$:
the initial velocities are $\dot \phi_0=0.379$ and
$\dot\psi_0=0.378$. The orbit differs in shape from the ones
seen above and its horizontal dimension is three times
smaller.  We observe eleven critical crossings between Region
I and Region II, corresponding to full angle reversals (middle
frame). There is no visible angle reversal (bottom frame).

We notice in each of the three cases that, regardless of the
apparent complexity of the orbits, when the system is near one
of the turning points ($\theta=\theta_X(t_j)$, red dots;
$\theta=\theta_N(t_j)$, blue dots), the points $(Q_R,Q_J)$ are
distributed along a straight line (top frame in each case).
In figure~\ref{fig:combined reversals (QR,QJ)} a plot is shown
combining the orbits of the three initial conditions used in
figures~\ref{fig:reversals(QR,QJ)1},
\ref{fig:reversals(QR,QJ)2} and \ref{fig:reversals(QR,QJ)3},
in order to compare their distribution and extent in the plane
$(Q_R,Q_J)$.

It is evident from the above that the criticality criterion is
a useful description of both full angle and visible angle
reversals. One just needs to initialize the system near a
critical line and the dynamics will do the rest. However, we
do not yet have an explanation for the extent of the orbit, so
our method is only descriptive and cannot predict, for
example, the number of rocking cycles executed in each
critical region. Forthcoming work should be dedicated to this
subject.

\begin{figure}
\begin{center}
\includegraphics[scale=0.21]{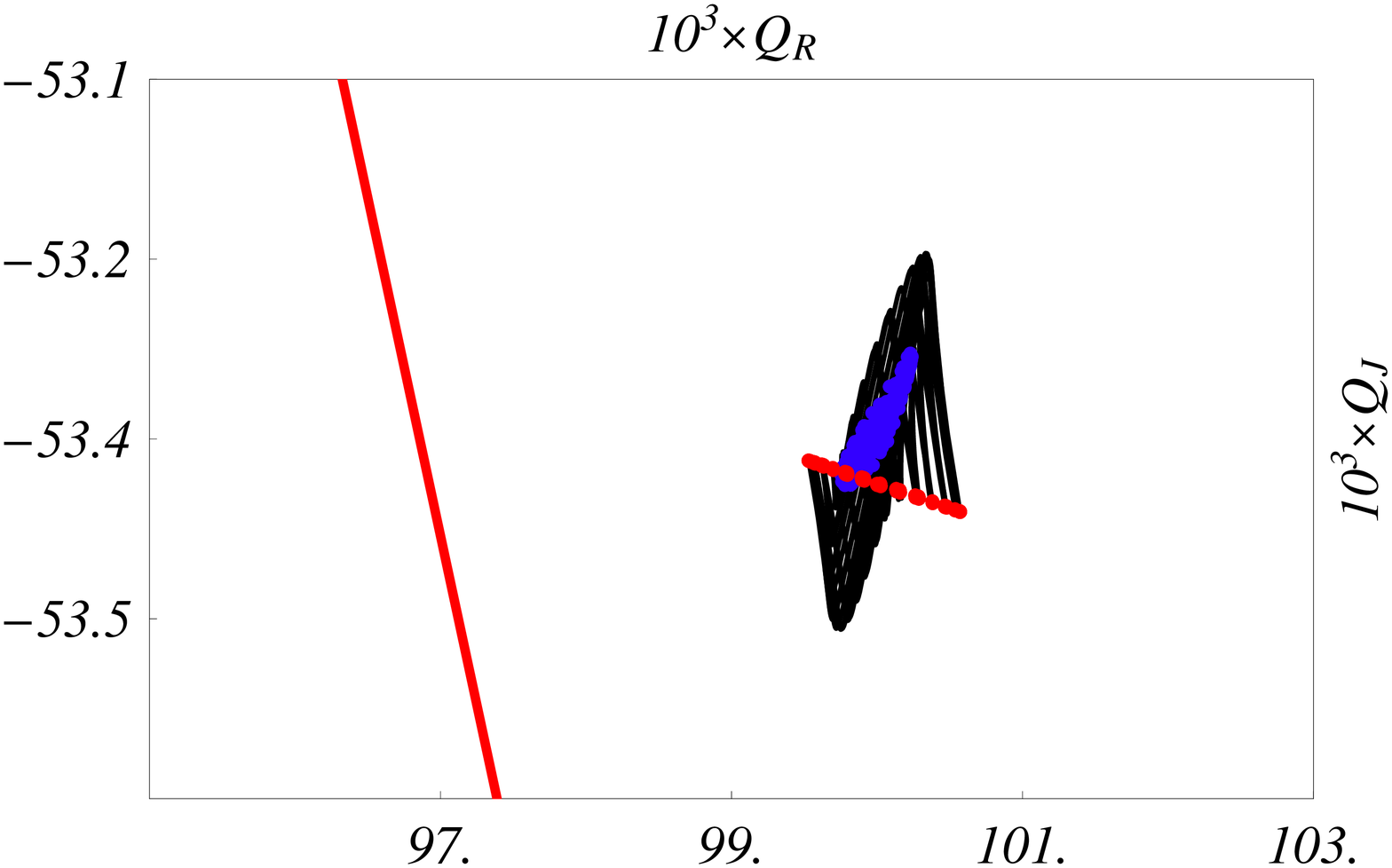}
\includegraphics[scale=0.21]{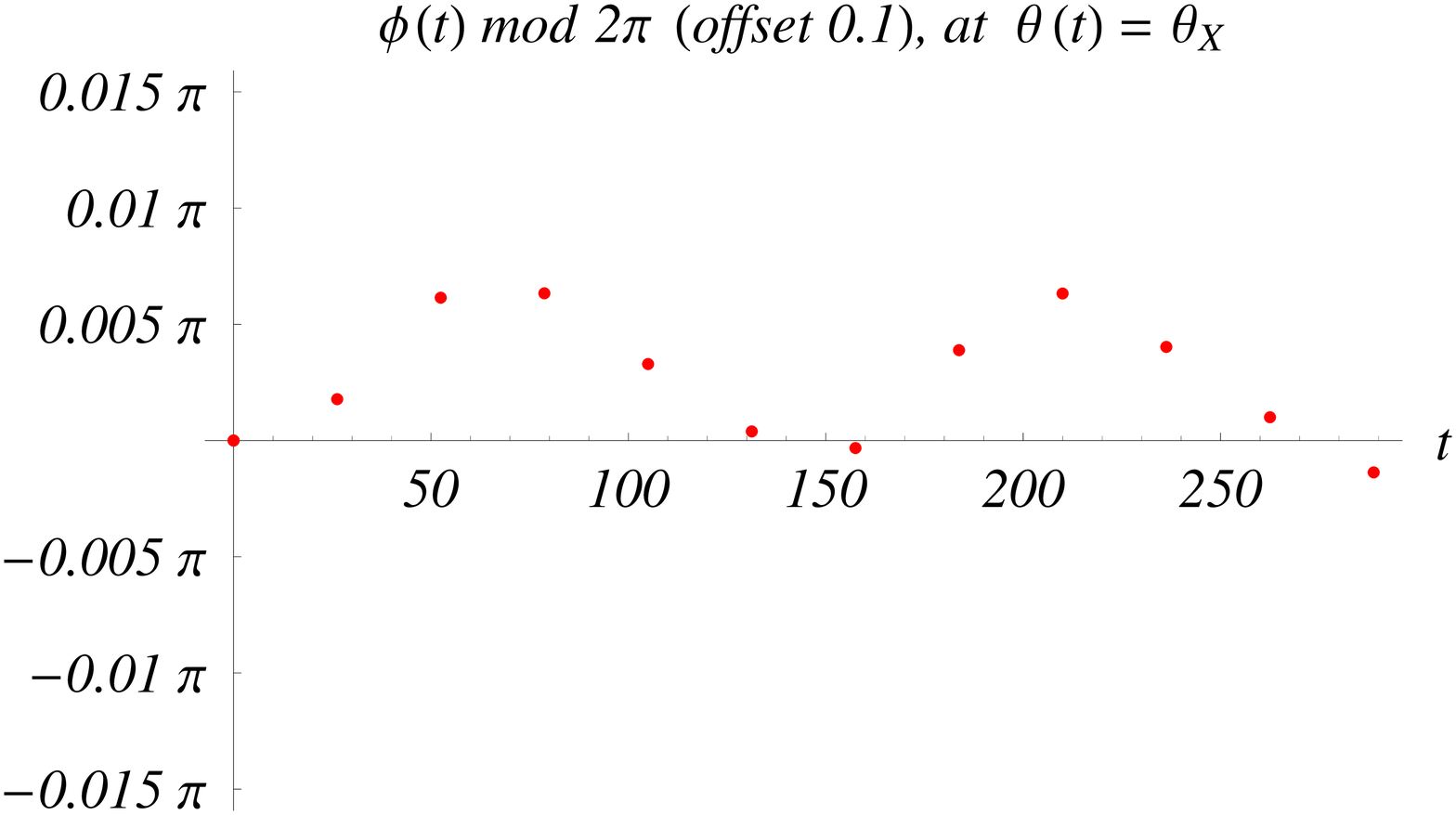}
\caption{(Colour online). \textbf{Left frame:} high-$Q_R$
reversal orbit, predicted by the asymptotic theory. Color code
as in figure \ref{fig:reversals(QR,QJ)1}. \textbf{Right
frame:} visible angle $\phi \, \mathrm{mod} \, 2 \pi,$ sampled
at times $t_j$ when $\theta(t_j)$ attains its maximum value
$\theta_X(t_j).$} \label{fig:high-Q_R}
\end{center}
\end{figure}

Regarding visible angle reversals, we have found that these
cease to be observed if the initial $(Q_R,Q_J)$ is chosen far
enough from the origin (keeping all other initial conditions
fixed). This can be understood from the fact that the
asymptotic Laurent expansion in (\ref{eq:phidotCJQR th=pi}) is
valid only near $\theta=\pi$ but the maximum attainable
$\theta(t)$ is bounded, from energy conservation, by
$\theta_0$, which is strictly less than $\pi$.
 From the analysis given in \S\ref{sec:range_X}, and equation
(\ref{eq:range_X}), it follows that the necessary condition
for validity of the Laurent expansion becomes $Q_R \leq 0.1$
for the present choice of parameters and initial conditions,
where we have used the observational estimate (from
figure~\ref{fig:combined reversals (QR,QJ)}) of $0.001$ for
the orbit extension along the $Q_R$-axis. We have checked that
there is indeed reversal for $Q_R=0.1$ (see
figure~\ref{fig:high-Q_R}, left frame). It is important to
mention that at this relatively high value of $Q_R$ (and
correspondingly high angular velocity) the lowest-order
Laurent asymptotic expansion given in (\ref{eq:phidotCJQR
th=pi}) needs to be improved. As a result, the simple
interpretation of reversals in terms of critical crossings and
changes of sign of $\dot\phi$ will change slightly. In
practice, to observe recession in this limiting case, it is
necessary to offset slightly the initial condition in the
plane $(Q_R, Q_J)$, to a point above the critical line
$Q_J=Q_{J,\pi}^{\mathrm{crit}}$. The resulting orbit remains
in Region I so that there is no change in sign of
$\dot\phi(t_j)$. However, the visible precession angle
$\Delta\phi(t_j)\,(\mathrm{mod}\,2\pi)$, being determined by
an integral in time, can and does have reversals
(figure~\ref{fig:high-Q_R}, right frame).

\begin{figure}
\begin{center}
\includegraphics[scale=0.3]{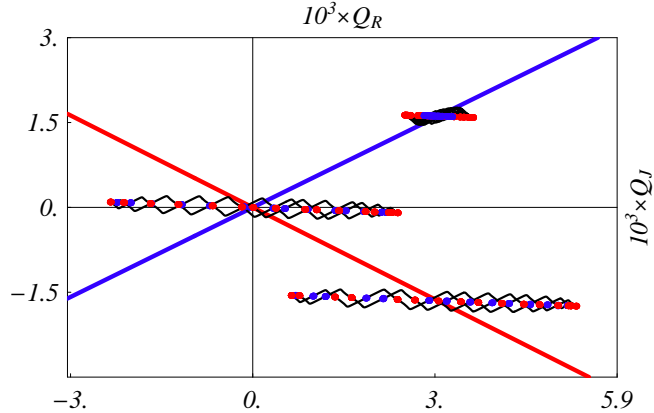}
\caption{(Colour online). Combined plot of orbits for the
three initial conditions described in
figures~\ref{fig:reversals(QR,QJ)1},
\ref{fig:reversals(QR,QJ)2} and \ref{fig:reversals(QR,QJ)3}.}
\label{fig:combined reversals (QR,QJ)}
\end{center}
\end{figure}

\section{Conclusion}
\label{sec:Conclusion}

Experiments show that the recession, or reversal of
precession, is a robust feature of the motion of the physical
\RnR. Analysis has confirmed that for a perfectly symmetric
body with $\IA=\IB$ this behaviour is impossible. However,
even the slightest breaking of this inertial symmetry is
sufficient to change the character of the solution profoundly,
allowing entirely new types of motion. Physical experiments
and numerical results show that the reversal angle
$\Phi=\phi_{\rm max}-\phi_{\rm min}$ depends sensitively on
the initial conditions. For motion that is initially close to
pure rocking, the angle $\Phi$ can be controlled by the choice
of the initial phase angle $\psi_0$. A rigourous analytical
demonstration of this result is outstanding.

The symmetric equations are integrable, with three invariants:
the total energy $E$, Jellett's quantity $Q_J$ and Routh's
quantity $Q_R$.  In the asymmetric case, only one of the above
three quantities is conserved, namely the total energy. We
present an analysis of recession based on the existence of
critical lines dividing the $(Q_R, Q_J)$-plane into 4
dynamically disjoint regions.  We prove that recession is
directly related to the lack of conservation of Jellett's and
Routh's quantities, by identifying individual reversals as
crossings of the orbit $(Q_R(t), Q_J(t))$ through the critical
lines. The criticality criterion allows one to produce a
family of initial conditions so that the system will exhibit
recession.


In the asymmetric case, there remains an underlying geometric
symmetry --- invariance under change of the azimuthal angle
$\phi$ --- so it is arguable that another dynamical invariant
exists. In the realistic case where gravity is present, this
additional integral (if it exists) remains to be found. Notice
that Borisov and Mamaev \cite{BorisovMamaev02} indicate in
their Table~1 that the quantity ${\bi M}^2 -2K{\bi r}^2$ is
conserved in the asymmetric case (where ${\bi M}$ is the
angular momentum about the contact point); however this is
only true in the absence of gravity.

There is apparently a slow period of the orbit $(Q_R(t),
Q_J(t))$. This suggests multi-scale analysis as an appropriate
technique for analysis of this problem. For small asymmetry
$\epsilon\equiv(\IB-\IA)/\IA$ the problem may be formulated as
a perturbed integrable Lagrangian system, and is amenable to
standard asymptotic analysis. This will be the subject of
future work. KAM theory \cite{Berry78, Ott95} would indicate
that certain aspects of integrability should apply to the
weakly asymmetric \RnR. However, the question of the general
integrability of the system remains open.


\section*{Acknowledgements}

We are grateful to Darryl Holm for inspiring conversations
during the course of this work.  We thank Brian O'Connor,
Senior Technical Officer in the UCD School of Physics, for
constructing a \RnR\ from a bowling ball.



\appendix
\section*{Appendix A}
\setcounter{section}{1}

\label{sec:appendix}


\subsection*{A.1: The basic parameters of the \RnR}

Let us assume that the body consists of homogeneous material
of uniform density, and that its mass and radius are both
unity.  We denote by $\Theta$ the co-latitude of the polar cap
that is removed to construct the \RnR. All the dynamical
parameters are determined once this angle is fixed. We define
the distance from the geometric centre to the centre of the
planar face of the body:
\[
  d = \cos\Theta
\]
The volume of the body is then
\[
  V = \pi \left( \frac{2}{3}+d-\frac{1}{3}d^3 \right)
\]
The off-set of the mass centre from the geometric centre is
\[
  a = \frac{\pi}{4} \left( \frac{d^2(2-d^2)-1}{V} \right)
\]
The moments of inertia about the geometric centre are
\[
\IC^\prime = \displaystyle{ \frac{\pi}{2} \left(
\frac{\frac{8}{15}+d-\frac{2}{3}d^3 +\frac{1}{5}d^5} {V}
\right) } \,, \quad \IA^\prime = \displaystyle{\pi\left(
   \frac{\frac{4}{15}+\frac{1}{4}d+\frac{1}{6}d^3
   -\frac{3}{20}d^5}{V} \right) }
\]
By means of the parallel axis theorem \cite{SyngeGriffith59},
the moments of inertia about the centre of mass are
\[
\IC = \IC^\prime \qquad \IA = \IA^\prime - a^2
\]
For the actual \RnR\ shown in figure~\ref{fig:bowlingball},
the polar angle is $\Theta\approx 53^\circ$. Thus $d = 0.6$,
giving the (nondimensional) parameter values
\[
 a = 0.085 \qquad \IA = 0.362 \qquad \IC = 0.42
\]
For our numerical experiments we used the values $a = 0.05$,
$\IA = 0.35$ and $\IC = 0.4$.

\subsection*{A.2: The equations for the symmetric \RnR}

The equations for the symmetric case $\IA=\IB$ were given in
\S\ref{sec:intermediateframe}. The details are given here.
Taking the cross-product of $\bi{r}$ with the momentum
equation (\ref{eq:vrot}) gives
\begin{equation}
\bi{r\cross}\bvdot + \bi{r\cross}(\bOmega\bi{\cross v})
  = \bi{r\cross F} = \bi{r\cross W} - \bi{G}
\label{eq:Gone}
\end{equation}
Noting that $\dot\theta=\omega_1$, the acceleration in
$\bi{\iprime\jprime\kprime}$-components is
\[
\bvdot = (f\dot\omega_2^\prime - s\dot\omega_3^\prime,
-f\dot\omega_1^\prime, s\dot\omega_1^\prime )
              + (-\omega_1^\prime(s\omega_2^\prime+c\omega_3^\prime),
s\omega_1^{\prime 2}, c\omega_1^{\prime 2}) \,.
\]
It follows that
\begin{eqnarray*}
\bi{r\cross}\bvdot = &&[(s^2+f^2)\dot\omega_1^\prime,
f^2\dot\omega_2^\prime-fs\dot\omega_3^\prime,
-fs\dot\omega_2^\prime+s^2\dot\omega_3^\prime)  \\
&& + (as\omega_1^{\prime 2},
-f\omega_1^\prime(s\omega_2^\prime+c\omega_3^\prime),
s\omega_1^\prime(s\omega_2^\prime+c\omega_3^\prime)]
\end{eqnarray*}
and
\begin{eqnarray*}
\bi{r\cross}(\bOmega\bi{\cross v})r
&=& \bi{(r\cdot v)}\bOmega - (\bi{r\cdot}\bOmega)v  \\
&=& -(
s^2+cf)(\omega_2^\prime/s)(f\omega_2^\prime-s\omega_3^\prime,
-f\omega_1^\prime, s\omega_1^\prime)
\end{eqnarray*}
Moreover,
\[
\bi{r\cross W} = - gas\,\bi{i^\prime}
\]
Using these expressions in (\ref{eq:Gone}) we get
\begin{eqnarray*}
G_1 &=& G_1^0 - (s^2+f^2)\dot\omega_1^\prime \\
G_2 &=& G_2^0 - (f^2\dot\omega_2^\prime-fs\dot\omega_3^\prime) \\
G_3 &=& G_3^0 - (s^2\dot\omega_3^\prime-fs\dot\omega_2^\prime)
\end{eqnarray*}
where, defining the height of the centre of mass as $h=1-ac$,
\begin{eqnarray*}
G_1^0 &=& -[as\omega_1^{\prime 2} -
           h(f\omega_2^\prime-s\omega_3^\prime)\omega_2^\prime/s] - gas \\
G_2^0 &=&
-[-f\omega_1^\prime(s\omega_2^\prime+c\omega_3^\prime) +
           hf\omega_1^\prime\omega_2^\prime/s] \\
G_3^0 &=&
-[s\omega_1^\prime(s\omega_2^\prime+c\omega_3^\prime) -
           h\omega_1^\prime\omega_2^\prime]
\end{eqnarray*}
We can now substitute for $\bi{G}$ in (\ref{eq:Rotation}) to
obtain
\begin{eqnarray}
[\IA+(s^2+f^2)]\dot\omega_1^\prime &= -
(\IC\Omega_2\omega_3^\prime-\IA\Omega_3\omega_2^\prime) &+
G_1^0
      \equiv P_1  \cr 
[\IA+f^2]\dot\omega_2^\prime + [-fs]\dot\omega_3^\prime &= -
(\IA\Omega_3\omega_1^\prime-\IC\Omega_1\omega_3^\prime) &+
G_2^0
      \equiv P_2
\label{eq:PPP}     \cr [-fs]\dot\omega_2^\prime +
[\IC+s^2]\dot\omega_3^\prime &=
&+ G_3^0
      \equiv P_3  
\end{eqnarray}
The first equation immediately gives the evolution of
$\omega_1^\prime$:
\[
\dot\omega_1^\prime = \frac{P_1}{\IA+s^2+f^2} \equiv S_1 \,.
\]
The second and third equations can be written
\[
\left[\matrix{   \IA+f^2  &   -fs  \cr
                  -fs    &  \IC+s^2  } \right]
\pmatrix{\dot\omega_2^\prime \cr \dot\omega_3^\prime} =
\pmatrix{  P_2  \cr P_3}
\]
The matrix is nonsingular, with determinant $\Delta =
(\IA\IC+\IA s^2+\IC f^2)$ and inverse
\[
\pmatrix{\dot\omega_2^\prime \cr \dot\omega_3^\prime} =
\frac{1}{\Delta}\left[\matrix{  \IC+s^2 &   fs  \cr
                                  fs   &  \IA+f^2  }\right]
\pmatrix{  P_2  \cr P_3} \equiv \pmatrix{  S_2  \cr S_3}
\]
The complete system of equations for the angular variables is
now obtained:
 \begin{eqnarray*}
 \dot\theta = \omega_1^\prime   \,, \qquad &&
 \dot\phi  = \omega_2^\prime/s   \,, \qquad
 \dot\psi = \omega_3^\prime - (c/s) \omega_2^\prime  \,. \\
 \dot\omega_1^\prime = S_1 \,, \qquad &&
 \dot\omega_2^\prime = S_2 \,, \qquad \dot\omega_3^\prime = S_3 \,. \
 \end{eqnarray*}
This system provides six equations for the six variables
$\{\theta,\phi,\psi,\omega_1^\prime,\omega_2^\prime,\omega_3^\prime\}$.


\subsection*{A.3: The Euler-Lagrange equations}

The Lagrange equations arising from (\ref{eq:Lagrangian}) may
be written
\begin{equation}
\bi{M}{\bthddot} + {\bi{P}}_{\btheta}(\btheta,\bthdot) =
\mathbf{0} \label{eq:ELtheta}
\end{equation}
where $\bthddot = (\ddot\theta,\ddot\phi,\ddot\psi)^{\rm T}$.
The symmetric matrix $\bi{M}$ is defined as
\[
\fl \bi{M} = \left[ \matrix{ \IA\chi^2+\IB\sigma^2+a^2s^2+h^2
& (\IA-\IB)s\sigma\chi              &     0     \cr
  (\IA-\IB)s\sigma\chi   & (\IA\sigma^2+\IB\chi^2+a^2)s^2+\IC c^2 & \IC c+as^2  \cr
            0             &   \IC c +as^2       &  \IC+s^2  }
\right]
\]
and the vector ${\bi{P}}_{\btheta} = (P_\theta,P_\phi,P_\psi)$
has components:
\begin{eqnarray*}
\fl {{P}}_{\theta} = [as]\dot\theta^2 +
[-(\IA\sigma^2+\IB\chi^2-\IC)sc + has]\dot\phi^2 +
[-(\IA-\IB)2\sigma\chi]\dot\theta\dot\psi  \cr +
[(\IA-\IB)s(\chi^2-\sigma^2)+\IC s+hs]\dot\phi\dot\psi + gas
\,, \cr \fl {{P}}_{\phi} = [(\IA-\IB)c\sigma\chi]\dot\theta^2
+ [(\IA\sigma^2+\IB\chi^2-\IC)2sc+(2ac-1)as]\dot\theta\dot\phi
\cr + [(\IA-\IB)s(\chi^2-\sigma^2)-\IC s +
asc]\dot\theta\dot\psi + [(\IA-\IB)2
s^2\sigma\chi]\dot\phi\dot\psi \,, \cr \fl {{P}}_{\psi} =
[(\IA-\IB)\sigma\chi]\dot\theta^2 +
   [-(\IA-\IB)s^2\sigma\chi]\dot\phi^2 \cr
+
[-((\IA-\IB)(\chi^2-\sigma^2)+\IC)s+(2ac-1)s]\dot\theta\dot\phi
+ [sc] \dot\theta\dot\psi   \,.
\end{eqnarray*}
Now (\ref{eq:ELtheta}) may be solved for
($\btheta(t),\bthdot(t))$. It has been confirmed, using {\em
Mathematica}, that the system (\ref{eq:thtddot}) is completely
equivalent to the system (\ref{eq:thetaomega}).



\end{document}